\documentclass[useAMS,usenatbib]{mn2e}

\usepackage{graphicx}
\usepackage{txfonts}
\usepackage{natbib}
\usepackage{xcolor}
\usepackage{siunitx,wasysym}
\usepackage{url}
\usepackage{longtable}

\def\teff{$T_{\rm eff}$}
\def\feh{[Fe/H]}
\def\logg{$\log g$}
\newcommand\mix[1]{\textsf{\textmd{\small{#1}}}}
\def\aj{AJ}
\def\araa{ARA\&A}
\def\apj{ApJ}
\def\apjl{ApJ}
\def\apjs{ApJS}
\def\apss{Ap\&SS}
\def\aap{A\&A}
\def\aapr{A\&A~Rev.}
\def\aaps{A\&AS}
\def\mnras{MNRAS}
\def\pasp{PASP}

\def\solphys{Solar Physics}
\def\ssr{Space Science Reviews}

\title[A new model spectral stellar library]{A new library of theoretical stellar spectra \\with scaled-solar and $\alpha$-enhanced mixtures}
\author[P. Coelho]{Coelho, P. R. T.$^{1,2}$ \thanks{pcoelho@usp.br}\\
$^{1}$Instituto de Astronomia, Geof\'isica e Ci\^encias Atmosf\'ericas, Universidade de S\~ao Paulo, 
R. do Mat\~ao 1226, 05508-090, S\~ao Paulo, Brasil\\
$^{2}$N\'ucleo de Astrof\'{\i}sica Te\'orica, Universidade Cruzeiro do Sul, 
R. Galv\~ao Bueno 868, 01506-000, S\~ao Paulo, Brasil\\
}

\begin{document}

\date{}

\pagerange{\pageref{firstpage}--\pageref{lastpage}} \pubyear{2002}

\maketitle

\label{firstpage}

\begin{abstract}
Theoretical stellar libraries have been increasingly used to overcome limitations of empirical libraries, e.g. by exploring atmospheric parameter spaces not well represented in the latter.
This work presents a new theoretical library which covers 3000 $\leq$ {\teff} $\leq$ 25000\,K, -0.5 $\leq$ {\logg} $\leq$ 5.5, and 12 chemical mixtures covering 0.0017 $\leq$ Z $\leq$ 0.049 at both scaled-solar and $\alpha$-enhanced compositions. 
This library complements previous ones by providing: \emph{i}) homogeneous computations of opacity distribution functions, models atmospheres, statistical surface fluxes and high resolution spectra; \emph{ii}) high resolution spectra with continua slopes corrected by the effect of predicted lines, and; \emph{iii}) two families of $\alpha$-enhanced mixtures for each scaled-solar iron abundance, to allow studies of the $\alpha$-enhancement both at `fixed iron' and `fixed Z' cases.
Comparisons to observed spectra were performed and confirm that the synthetic spectra reproduce well the observations, although there are wavelength regions which should be still improved. The atmospheric parameter scale of the model library was compared to one derived from a widely used empirical library, and no
systematic difference between the scales was found. This is particularly reassuring for methods which use synthetic spectra for deriving atmospheric parameters of stars in spectroscopic surveys. 
\end{abstract}


\section{Introduction}
\label{s:intro}

Libraries of stellar spectra are important in a variety of areas:  
\emph{(a)} deriving atmospheric parameters in stellar surveys, via automatic analysis and classification of data;
\emph{(b)} determination of radial velocities via cross-correlation against templates, e.g., for the detection of exoplanets;
\emph{(c)} calibration of features for spectroscopic classification;
\emph{(d)} calibration of photometric indices, and;
\emph{(e)} in the study of the star formation history of galaxies as a core ingredient to stellar population models. 

A stellar library is at the heart of accurate stellar population models, and 
should ideally provide complete coverage of the HR diagram, accurate atmospheric parameters 
(effective temperature \teff, surface gravities {\logg} and abundances [Fe/H], [Mg/Fe], etc.), good compromise between wavelength coverage, spectral resolution and signal-to-noise (S/N).
Both empirical and theoretical libraries can be used for this purpose, and which choice is ``best'' is a matter of on-going debate at related conferences and literature \citep[see e.g.][and references therein]{coelho09a_proc}.

The main caveat of empirical libraries is the limited coverage of the HR diagram: hot stars are not well sampled and abundance patterns are biased towards the solar neighbourhood. With the advent of modern extragalactic surveys, this limitation hampered our ability of studying stellar populations which have undergone a star formation history very different from the one in our vicinity. The first compelling evidence of this limitation was presented by \citet{worthey+92}, who showed that stellar population models for Lick/IDS indices cannot 
reproduce the indices measured in elliptical galaxies, indicating that these systems are 
overabundant in $\alpha$-elements relative to the Sun. This is a direct consequence of the fact that, by construction, 
the abundance pattern of stellar population models based on empirical libraries is dictated by that of the 
library stars, which is dominated by the abundance pattern of the solar neighbourhood \citep[e.g.][]{mcwilliam97}.

Theoretical libraries can be used to overcome this limitation and several are available in literature \citep[e.g.][sampling only the last decade]{barbuy+03,murphy_meiksin04,zwitter+04,coelho+05,martins+05,munari+05,UVBLUE05,fremaux+06,leitherer+10,palacios+10,sordo+10,kirby11,delaverny+12}. 
Moreover, a theoretical stellar spectrum has very well defined atmospheric parameters, does not suffer from low S/N, and covers a larger wavelength range at a higher spectral resolution than any observed spectrum. 

On the other hand, being based on our knowledge of the physics of stellar atmospheres and databases of atomic and molecular opacities, theoretical libraries are limited by the approximations and (in)accuracies of their underlying models and input data \citep[e.g.][]{bessell+98,kucinskas+05,kurucz06_lines,MC07,bertone+08,coelho09a_proc,plez11proc,lebzelter+12,sansom+13}.

Besides, a theoretical spectral library that is intended to reproduce
high resolution spectra is not a library that also
predicts good spectrophotometry. That happens because when computing
a synthetic spectrum, one has to choose to include or not the
so-called `predicted lines': lines where either one or both energy levels of the transition were predicted from quantum mechanics calculations \citep[][]{kurucz92}, as opposed to lines whose energy levels were measured in laboratory. Usually only the lower
energy levels of atoms have been determined in the laboratory, particularly for complex atoms such as iron. If only those transitions were taken into account, the atmospheric line blanketing computed from such data would be severely incomplete. The predicted lines are essential
for computing accurately the structure of model atmospheres and for spectrophotometric predictions \citep[e.g.][]{short_lester96}. But as the quantum mechanics
predictions are accurate to only a few per cent, wavelengths for these lines may be largely uncertain, and the line oscillator strengths are sufficiently accurate merely in a statistical sense \citep{kurucz06_lines}. The
predicted lines are, therefore, unsuitable for high resolution analyses \citep{bell+94,castelli_kurucz04,munari+05}.
In practice, theoretical libraries 
aimed at spectrophotometric calibrations include the predicted lines, 
while libraries aimed at high resolution studies are 
computed with shorter, fine-tuned, often empirically calibrated atomic and molecular line lists 
\citep[e.g.][]{peterson+01,barbuy+03,coelho+05,UVBLUE05}.
With current atomic data available, either choice is only a compromise solution.

Despite these limitations, model stellar spectra opened new important ways to study integrated light from stellar populations. Theoretical stellar libraries have been used to build fully theoretical stellar population models \citep{leitherer+99,delgado+05,coelho+07,percival+09,buzzoni+09proc,lee+09}, and were crucial for the development of methods which lead to the measuring of element abundances (beyond the global metal content) in integrated stellar populations \citep[e.g.][]{trager+98,proctor_sansom02,thomas+05}. In recent years, model spectra are flourishing in extragalactic applications by allowing the spectral modelling of a variety of stellar histories via differential methods, i.e., combining empirical stellar libraries with model predictions \citep[][]{cervantes+07proc,prugniel+07proc,walcher+09,conroy_dokkum12}.

This work is the first of a series aiming at expanding our stellar population modelling \citep{coelho+05,coelho+07,walcher+09} towards larger coverage in ages, metallicities and wavelength range. A library of theoretical stellar spectra is presented, bringing:
\emph{(a)} a homogeneous computation of opacity distribution functions, model atmospheres, statistical samples of surface fluxes from 130\,nm to 100\,$\mu$m for low resolution studies, and high resolution synthetic spectra computed from 250 to 900\,nm;
\emph{(b)} high resolution spectra with continuum slopes corrected for the effect of predicted lines, and;
\emph{(c)} two families of $\alpha$-enhanced mixtures for each scaled-solar iron abundance, to allow differential studies of the $\alpha$-enhancement both at `fixed iron' and `fixed Z' cases. 
The present library is not intended as a direct replacement for the one presented in \citet{coelho+05} as the later was tailored at the modelling of stars of spectral types G, K and early-M. The present library employed different codes and opacities and covers a larger range of effective temperatures, being favoured in differential spectral analysis.
Stellar population models built with the present library will be published in a forthcoming paper (Coelho et al. 2014, in prep).

Section \ref{s:lib} describes how the models were computed and the ingredients adopted. 
The effect of the predicted lines is discussed and quantified in \S\ref{s:pl}.
Section \S\ref{s:comparisons} compares the model predictions with observations: 
a comparison with an empirical colour-temperature calibration is given in \S\ref{s:colours};  
and in \S\ref{s:spectra}, \S\ref{s:atmpars} the model spectra are compared to an empirical spectral library.
Concluding remarks are given in \S\ref{conclusions}.

\section{The theoretical library of model atmospheres, fluxes and spectra}
\label{s:lib}

The library consists of opacity distribution functions, model atmospheres, statistical samples of surface fluxes (\mix{SED} models) and high resolution synthetic spectra (\mix{HIGHRES} models).
Each of these components of the library is explained in detail below and the whole library is publicly available to the astronomical community.

The \mix{SED} and \mix{HIGHRES} models can be retrieved from the Spanish Virtual Observatory\footnote{\url{http://svo.cab.inta-csic.es/main/index.php}} (SVO; \citealt{gutierrez+06proc}) and from the website of the author\footnote{\url{http://www.astro.iag.usp.br/}$\sim$\url{pcoelho/}}, as FITS files \citep{pence+10}. The files can be queried via web interface at the SVO Theoretical Data Server\footnote{\url{http://svo2.cab.inta-csic.es/theory/newov/}} or via any software that is compliant to the VO TSAP protocol\footnote{\url{http://svo2.cab.inta-csic.es/theory/docs2/index.php?pname=TSAP/How}\%\url{20To}}. The ODF and model atmosphere files can be obtained upon request to the author.

The target use of the present library, although not limited to that, is the spectral modelling of stellar populations and the measurement of ages, iron abundances and $\alpha$ over iron ratios in integrated light \citep{coelho+07,walcher+09}. As such, the coverage of the parameters {\teff} and {\logg} were fine-tuned to encompass evolutionary stages relevant to the integrated light of populations with ages between 30\,Myr and 14\,Gyr, from lower main sequence to the early Asymptotic Giant Branch.

The library encompasses 12 different chemical mixtures, summarised in Table \ref{tab_mixes}. These mixtures were chosen to be consistent with a new grid of stellar evolutionary tracks to be presented in a forthcoming paper on stellar population models (Coelho et al. 2014, in prep.).

\begin{table}
\caption{\label{tab_mixes} Chemical mixtures covered by the grid. The values in the last two columns are given relative to the solar mixture from \citet{gs98}. The usual spectroscopic notation for abundances is used: \feh\,= $\log(N_{\rm Fe}/N_{\rm H})_{\rm star} - \log(N_{\rm Fe}/N_{\rm H})_{\rm Sun}$, where N$_x$ is the number density of atoms of each elemental species.}
\begin{tabular}{lccccc}
\hline
Label & X & Y & Z & [Fe/H] & [$\alpha$/Fe]\\
\hline
\mix{m10p00}  &  0.7563 & 0.2420  & 0.0017   &        -1.0     &    0.0 \\
\mix{m13p04}  &  0.7563 & 0.2420  & 0.0017   &        -1.3     &    0.4 \\
\mix{m10p04}  &  0.7515 & 0.2450  & 0.0035   &        -1.0     &    0.4  \\
\hline
\mix{m05p00}  &  0.744  & 0.251  & 0.005    &        -0.5     &    0.0  \\
\mix{m08p04}  &  0.744  & 0.251  & 0.005    &        -0.8     &    0.4  \\
\mix{m05p04}  &  0.739  & 0.250  & 0.011    &        -0.5     &    0.4  \\
\hline
\mix{p00p00}  &  0.717  & 0.266  & 0.017    &         0.0     &    0.0  \\
\mix{m03p04}  &  0.717  & 0.266  & 0.017    &        -0.3     &    0.4  \\
\mix{p00p04}  &  0.679  & 0.289  & 0.032    &         0.0     &    0.4  \\
\hline
\mix{p02p00}  &  0.708  & 0.266  & 0.026    &         0.2     &    0.0  \\
\mix{m01p04}  &  0.708  & 0.266  & 0.026    &        -0.1     &    0.4  \\
\mix{p02p04}  &  0.642  & 0.309  & 0.049    &         0.2     &    0.4  \\
\hline
\end{tabular}
\end{table}

The mixtures consist of four scaled solar mixtures \citep{gs98} and eight $\alpha$-enhanced mixtures ([$\alpha$/Fe] = 0.4\,dex, where $\alpha$-elements are O, Ne, Mg, Si, S, Ca and Ti). Each of the scaled solar mixtures (\mix{m10p00}, \mix{m05p00}, \mix{p00p00}, \mix{p02p00}) has two corresponding $\alpha$-enhanced mixtures: 
\begin{itemize}
\item one where the iron abundance [Fe/H] was kept constant relative to the scaled solar counterpart, thus enhancing the metalicity Z (where Z is the mass fraction of metals; \mix{m10p04}, \mix{m05p04}, \mix{p00p04} and \mix{p02p04}), and; 
\item another where Z was kept constant, thus lowering [Fe/H] (\mix{m13p04}, \mix{m08p04}, \mix{m03p04}, \mix{m01p04}). These mixtures can also be understood as '\emph{iron poor}' patterns at constant Z. 
\end{itemize}

The motivation to compute two $\alpha$-enhanced mixtures for each scaled-solar mixture is that stellar evolution tracks are traditionally parametrized in terms of Z \citep[e.g.][]{pietrinferni+04}, while stellar spectral libraries are parametrized in terms of iron abundance [Fe/H] \citep[e.g.][]{MILES2}. The link between Z and [Fe/H] is not always straightforward, and some stellar population models are parametrized in terms of iron content \citep[e.g.][]{schiavon07,coelho+07} while others are parametrized in terms of total metal content \citep[e.g.][]{trager+98,BC03}. 

The values of [Fe/H] and [$\alpha$/Fe] reported in Table \ref{tab_mixes} are given adopting the solar abundance pattern by \citet{gs98}. With respect to the newer determinations of solar abundances, [Fe/H] is unchanged with \citet{asplund+09} and -0.02\,dex should be added to the reported [Fe/H] values for the \citet{caffau+11} scale. The conversion of [$\alpha$/Fe] to the newer solar patterns depends on the proxy $\alpha$-element of choice. For Oxygen, 0.14 and 0.07\,dex should be added to the reported [$\alpha$/Fe] values for \citet{asplund+09} and \citet{caffau+11} scales, respectively. For Magnesium, -0.02\,dex should be added to [$\alpha$/Fe] for \citet{asplund+09} scale (\citealt{caffau+11} do not provide determinations of Mg).

\begin{figure}
\begin{center}
\includegraphics[width=\columnwidth]{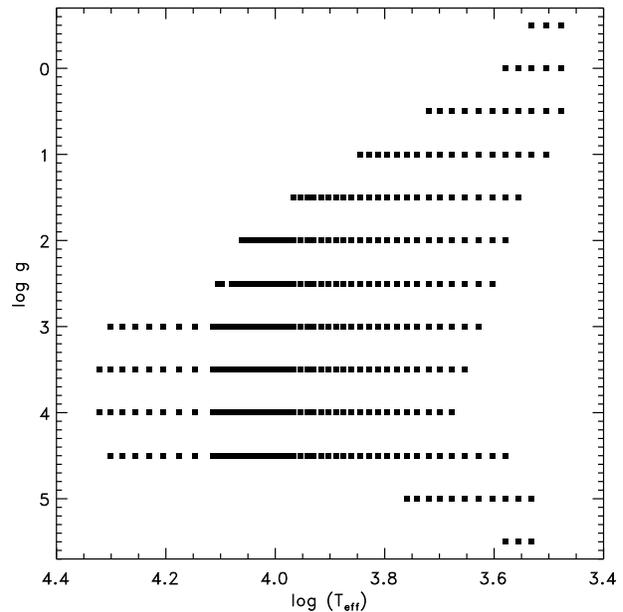}
\caption{The coverage of the stellar library in the plane {\teff} (x-axis) vs. {\logg} (y-axis), for the solar  mixture (\mix{p00p00} in Table \ref{tab_mixes}).}
\label{fig_coverage}
\end{center}
\end{figure}

For each mixture in Table \ref{tab_mixes}, the values for {\teff} and {\logg} were chosen to cover the \emph{loci} occupied by isochrones between 30\,Myr and 14\,Gyr (computed by A. Weiss, to be presented in Coelho et al. 2014, in prep.). The exact coverage is, therefore, slightly different from mixture to mixture, and ranges from {\teff} = 3000 to 25000\,K (in steps of 200\,K below {\teff} = 4000\,K, 1000\,K above {\teff} = 12000\,K and 250\,K otherwise), and {\logg} from -0.5 to 5.5\,dex (in steps of 0.5\,dex). The coverage of the mixture \mix{p00p00} (solar abundances) is presented in Fig. \ref{fig_coverage} in the plane log(\teff) vs. \logg, for illustration purposes. Each point in the figure has a correspondent model atmosphere, statistical flux distribution and high resolution spectrum.

\begin{figure*}
\begin{center}
\includegraphics[bb=20 30 450 450,width=1.5\columnwidth]{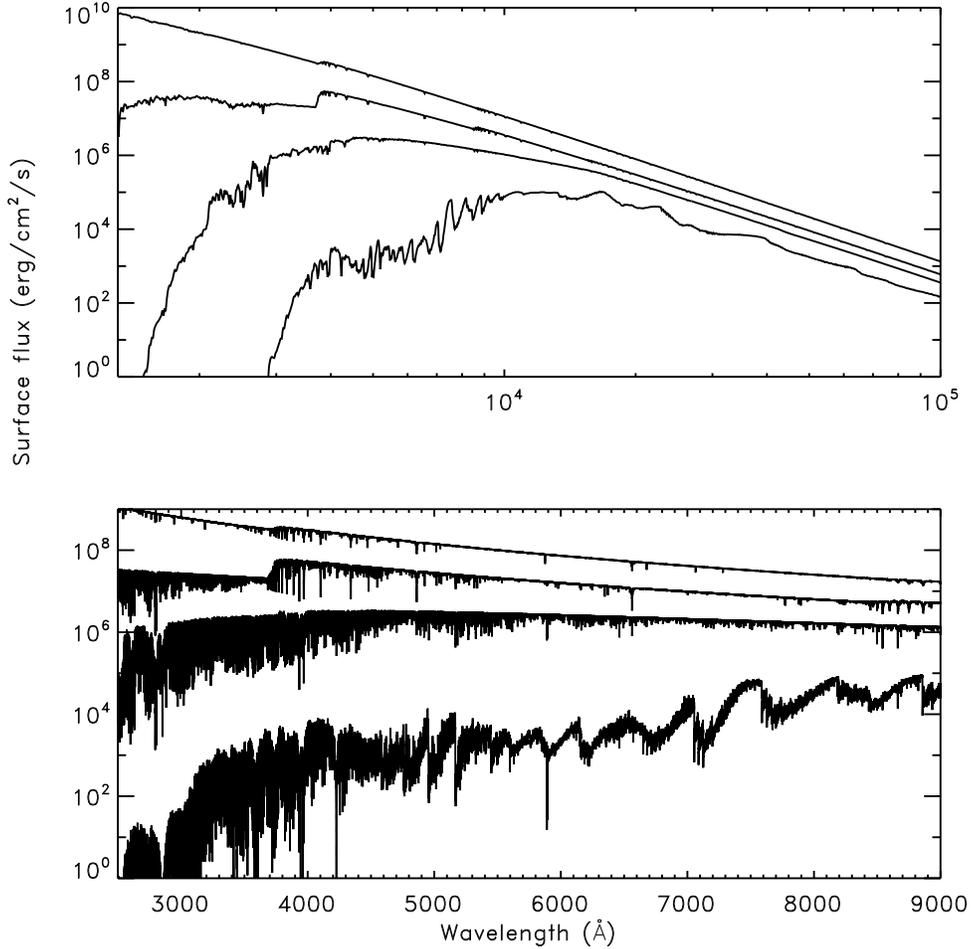}
\caption{\mix{SED} and \mix{HIGHRES} models are illustrated in the top and bottom panels, respectively. In both panels, the atmospheric parameters ({\teff}, {\logg}, [Fe/H] and [$\alpha$]/Fe]) of the models shown are: (26000\,K, +4.0\,dex, -1.3, +0.4), (10000\,K, +2.5\,dex, +0.2, +0.4), (5750\,K, +4.5\,dex, 0.0, 0.0) and (3000\,K, +0.0\,dex, -0.5, 0.0), from top to bottom. The full library is publicly available online (see text in \S\ref{s:lib}).}
\label{f:sample_library}
\end{center}
\end{figure*}

Figure \ref{f:sample_library} illustrates some of the spectral models available. The computation and characterisation of the library is fully described in the following sections.

\subsection{Opacity distribution functions (\mix{ODF}) and model atmospheres}
\label{s:odf}

It is convenient from the computational point of view to split the calculation of a theoretical spectra in two major steps: the calculation of the model atmosphere, commonly adopting opacity distribution function \citep[ODF;][]{strom_kurucz66} or opacity sampling techniques \citep[OS;][]{johnson_krupp76} – and the calculation of the spectrum with a spectral synthesis code.
The OS technique can directly produce as output a sampled flux distribution, but is more time consuming from the computational point of view. 

A model atmosphere gives the run of temperature, gas, electron and radiation pressure, convective velocity and flux, and more generally, of all relevant quantities as a function of some depth variable (geometrical, or optical depth at some special frequency, or column mass) in a stellar photosphere of given atmospheric parameters.

For the present library, ODF for all mixtures were computed with the Linux port of the code DFSYNTHE \citep{kurucz_codes05,kurucz05,DFSYNTHE}. Extensive grids of  ODFs had been computed recently in literature \citep{ATLASODFNEW, kirby11, meszaros+12}, but for mixtures different from the ones adopted in this work. 

Based on the newly computed ODFs, model atmospheres were computed using a linux port of the code ATLAS9 \citep{ATLAS1970,ATLAS_LINUX} for stars with {\teff} $\ge$ 4000\,K. 
Atmosphere models were computed under the assumption of plane-parallel geometry, using the turbulent velocity $\xi$  = 2\,km/s
and mixing length parameter $\alpha_{ML}$ = 1.25.
The convergence criteria for model atmosphere calculations are  similar to those adopted in \citet{meszaros+12}: no more than one non-converged layer was accepted between log $\tau_{Ross}$ = -- 5 and log $\tau_{Ross}$ = 1, where $\tau_{Ross}$ is the Rosseland optical depth (as most of the lines from the optical to the H band form in this interval). 90\% of the models have converged through the whole atmosphere. 

For stars below {\teff} = 4000\,K, pre-computed MARCS model atmospheres were adopted\footnote{Available at \url{http://marcs.astro.uu.se/}} \citep{MARCS08}, as these models are computed with a larger set of molecular opacities important to the atmosphere structure of cool stars (in particular VO and ZrO). Additionally, MARCS models for giants are computed at spherical symmetry, as at these very low effective temperatures the atmospheres of late-type giants become very extended, and thus stellar atmosphere models employing plane-parallel geometry (e.g., ATLAS) are not adequate. 
  
ATLAS codes \citep{kurucz_codes05} use atomic and molecular line lists made available by R. Kurucz through his website\footnote{\url{http://kurucz.harvard.edu/}}. The list comprises the molecules: C$_2$ (systems A-X, B-A, D-A, E-A); CH (A-X, B-X C-X); CN (A-X, B-X); CO (A-X, X-X); H$_2$ (B-X, C-X); MgH (A-B, B-X); NH (A-X, C-A); OH (A-X, X-X); SiH (A-X); SiO (A-X, E-X, X-X); TiO ($\alpha$, $\beta$, $\gamma$, $\gamma$', $\delta$, $\phi$, $\epsilon$), and; H$_2$O. The molecular line lists for TiO and H$_2$O are reformatted versions of the lists presented in \citet{schwenke98} and \citet{partridge_schwenke97}, respectively. The lists for SiH and OH were recomputed and published online in recent years by R. Kurucz, and the list of H$_2$O was also recently corrected. The other lists are the same as provided in \citet{kurucz93CD15}. 

\subsection{Spectral energy distributions (SED) at low resolution}
Statistical samples of model fluxes and synthetic spectra correspond to emergent flux predicted by a model atmosphere, and are required for comparison with observations.
Statistical samples of the model surface fluxes have been commonly used in literature as low resolution spectral energy distributions \mix{SEDs} \citep[e.g.][]{basel1,basel2,basel3}. 
These fluxes are associated with the model atmosphere computation, where the radiative-transfer equation is solved at a given number of frequency points, chosen to properly sample the spectral regions where the radiation field is strong. A detailed knowledge of the radiative field is not critical for stellar atmosphere models because their structural properties depend on global aspects of the radiation field \citep[e.g.][]{leblanc2010}. 
The sampled fluxes are thus adequate to compute synthetic broadband photometry only, while higher resolution synthetic spectra are needed for narrow-band photometry and spectroscopy \citep[see][]{MARCS08, plez08proc}.

Statistical samples of model fluxes in the present library were computed with the code ATLAS9v, made available by F. Castelli at her website\footnote{\url{http://wwwuser.oat.ts.astro.it/castelli/sources/atlas9codes.html}}. The opacities considered are the same ones used for the model atmosphere computations, described in \S\ref{s:odf}. The models are available as FITS files covering from 130\,nm to 100\,$\mu$m at a wavelength sampling of $\Delta \log\lambda = 8\times10^{-4}$. 
 For computing the \mix{SEDs}, opacities due to predicted lines are included, to ensure a better modelling of photometric properties (see discussion in \S\ref{s:intro} and \S\ref{s:pl}).

\subsection{High spectral resolution library (HIGHRES)}

High resolution synthetic spectra were computed from 250 to 900\,nm with the spectral synthesis code SYNTHE
\citep{kurucz_avrett81} in its public Linux port by \citet{ATLAS_LINUX}. 
This wavelength range includes several features largely used to measure iron and $\alpha$-elements abundances, from the magnesium triplet in the UV to the Calcium triplet in the near infrared, and also the Balmer jump and several Hydrogen lines largely used in age determinations of stellar populations.
The models were computed at a 
wavelength sampling\footnote{The term \emph{resolution} in the context of a model spectral library might lead to some confusion, as the word often indicates different concepts in the numerical modelling community and in the spectroscopy community. Models are computed at a given numerical resolution which defines the frequency points for the radiate transfer evaluation. This characterizes the wavelength sampling of the output model spectrum and sometimes is refereed to as '\emph{wavelength resolution}'. Prior to use, models are often broadened to a '\emph{spectral resolution}', simulating a specific line-spread-function such as an instrumental spectral resolution, rotational broadening or velocity dispersion. Both wavelength and spectral '\emph{resolutions}' can be parametrized in terms of R $= \lambda / \Delta\lambda$, but in the first case, $\Delta\lambda$ is the wavelength step while in the second case, $\Delta\lambda$ corresponds to the full-width half maximum of the line-spread function. The Nyquist-Shannon sampling theorem \citep{nyquist,shannon} tells us that the sampling frequency should be greater than twice the highest frequency contained in the signal. In the context of spectra, this translates that the spectral resolution is, at maximum, half the value of the wavelength resolution.} of R$_{\lambda}$ = 300000, 
broadened by a gaussian line-spread function of R$_{\rm LSF}$ = 20000 and resampled to a constant wavelength sampling of 0.02\,\AA. 
For higher spectral resolution analysis, the unbroadened spectra are available upon request to the author. The computed fluxes correspond to stellar surface fluxes in units of erg/cm$^2$/s/\AA. SYNTHE assumes plane-parallel models are provided as input, though some of the models for giants stars (the ones from MARCS) are computed in spherical symmetry. \citet{heiter_eriksson06} have shown that the effect on the line profiles due to this inconsistency is rather small, and that consistency seems to be less important than using a spherical model atmosphere, when appropriate. 

The atomic line list adopted is based on the compilations by
\citet{coelho+05} and \cite{castelli_hubrig04}, and the reader is refereed to those references for details on how the lines were calibrated. For lines in common between the two lists, atomic transition parameters (central wavelength, energy level, oscillator strength and broadening) that best reproduced the solar spectrum \citep{solarflux:kurucz84} were kept. \citet{MC07} have shown that \citet{coelho+05} library on average better reproduced spectral indices of F, G and K stars, when compared to \citet{martins+05} and \citet{munari+05} libraries, due to its line list calibration. But complementing with the \cite{castelli_hubrig04} line list was important for the higher ionisation metal lines and Paschen H lines, not included in \citet{coelho+05}. Atomic lines with predicted energy levels were not included in the \mix{HIGHRES} models, for the reasons explained in \S\ref{s:intro} and \S\ref{s:pl}. 

Lines for the molecules C$_2$, CH, CN,
CO, H$_2$, MgH, NH, OH, SiO and SiH were included for all stars, and  
TiO lines were included for stars cooler than {\teff} = 4500\,K, from the sources described in \S\ref{s:odf}.
The lack of VO in the molecular line list prevented the calculation of stars with spectral type later than M7 \citep{tsuji86}, which correspond to stars with {\teff} around 2800 -- 3200 \citep[][]{dyck+96,kucinskas+05,rajpurohit+13}. Besides, there is evidence of dust forming in the upper layers of stars with {\teff} below 3000\,K, veiling the visual flux \citep[e.g.][]{jones_tsuji97}. Therefore, the coolest star computed in each mixture was set to {\teff} = 3000\,K.

\section{Effect of predicted lines in the fluxes: evaluation and correction}
\label{s:pl}

The effect of the atomic lines with predicted energy levels (`\emph{predicted lines}', PLs) in high spectral resolution features has been discussed and shown in e.g. \citet{munari+05}. 
The authors compared synthetic spectra computed with and without predicted lines with observations, and addressed the effect of the PLs on cross-correlation determination of radial velocities and analysis of binary components. They show that there are wavelength intervals where strong PLs cluster together `polluting' the model spectrum with unobserved lines (lines whose central wavelengths and/or oscillator strengths are severely wrong; see Fig. 3 in \citealt{munari+05} and Fig. 10 in \citealt{bell+94}). This also results in radial velocities determination significantly worse when model templates were adopted from a library computed with PLs. Their conclusion holds true for the present library, as no significant improvement has been made to the PL list since then. Progress is expected in the near future (R. Peterson, priv. comm.).  

\begin{figure}
\begin{center}
\includegraphics[width=\columnwidth]{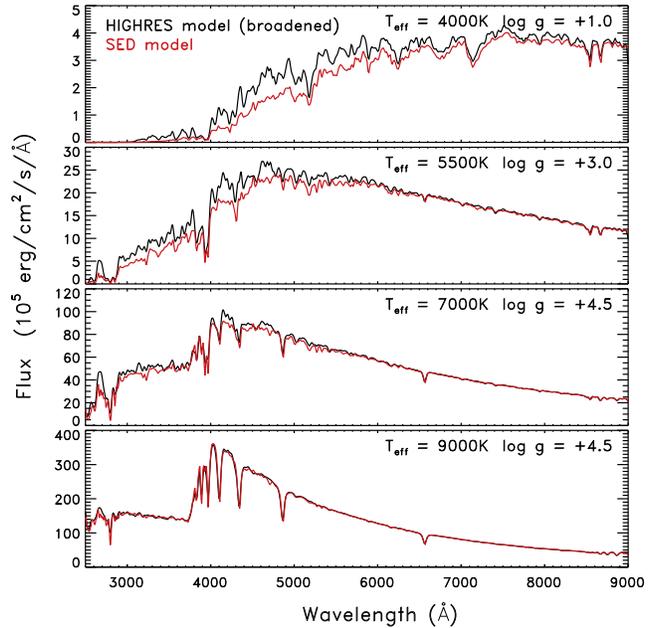}
\caption{\mix{SED} and \mix{HIGHRES} models (red and black lines respectively) are compared for four combinations of stellar parameters, indicated in each panel. All models have super-solar abundances (\mix{p02p04}) and were broadened to a resolution of FWHM $\sim$ 30\,{\AA} for easier visualization. The blanketing due to the inclusion of PLs is easily noticed in the coolest star and diminishes as temperature increases.}
\label{fig_plexample}
\end{center}
\end{figure}

On the other hand, the lack of PLs underestimate the blanketing (mostly in the blue bands), affecting the predictions of broad band colours. 
\citet{coelho+07} have shown that stellar population models based on a library without the PLs understimate the U--B colour of simple stellar populations by more than 0.2\,mag, and the B--V colour by $\sim$ 0.1\,mag. In order to provide good predictions for 
both high spectral resolution features and broad-band colours in stellar population models, either libraries that do not 
include the predicted lines must be `flux calibrated' \citep[section 3.2 in][]{coelho+07} or low and high
resolution stellar population models should be computed with different libraries \citep[as adopted by e.g.][]{percival+09}.

In this section, the photometric effect of the PL is quantified by the comparison between the \mix{SED} and \mix{HIGHRES} libraries. The goal is to obtain smooth flux corrections to be applied to the \mix{HIGHRES} models in order to make them suitable to stellar population modelling of both photometric and spectroscopic features. Besides, `flux calibrated' \mix{HIGHRES} models are more reliable in techniques of spectral fitting which take into account the continuum slope, such as the ones performed with the code {\sc Starlight} \citep{cid+05}.

The blanketing due to the PLs affects mostly the blue part of the spectra (becoming progressively fainter with larger wavelengths) and varies with the atmospheric parameters. 
Fig. \ref{fig_plexample} shows comparisons for four combinations of atmospheric parameters, selected to illustrate the dependence of the PL blanketing with temperature.

\begin{figure}
\begin{center}
\includegraphics[width=0.9\columnwidth]{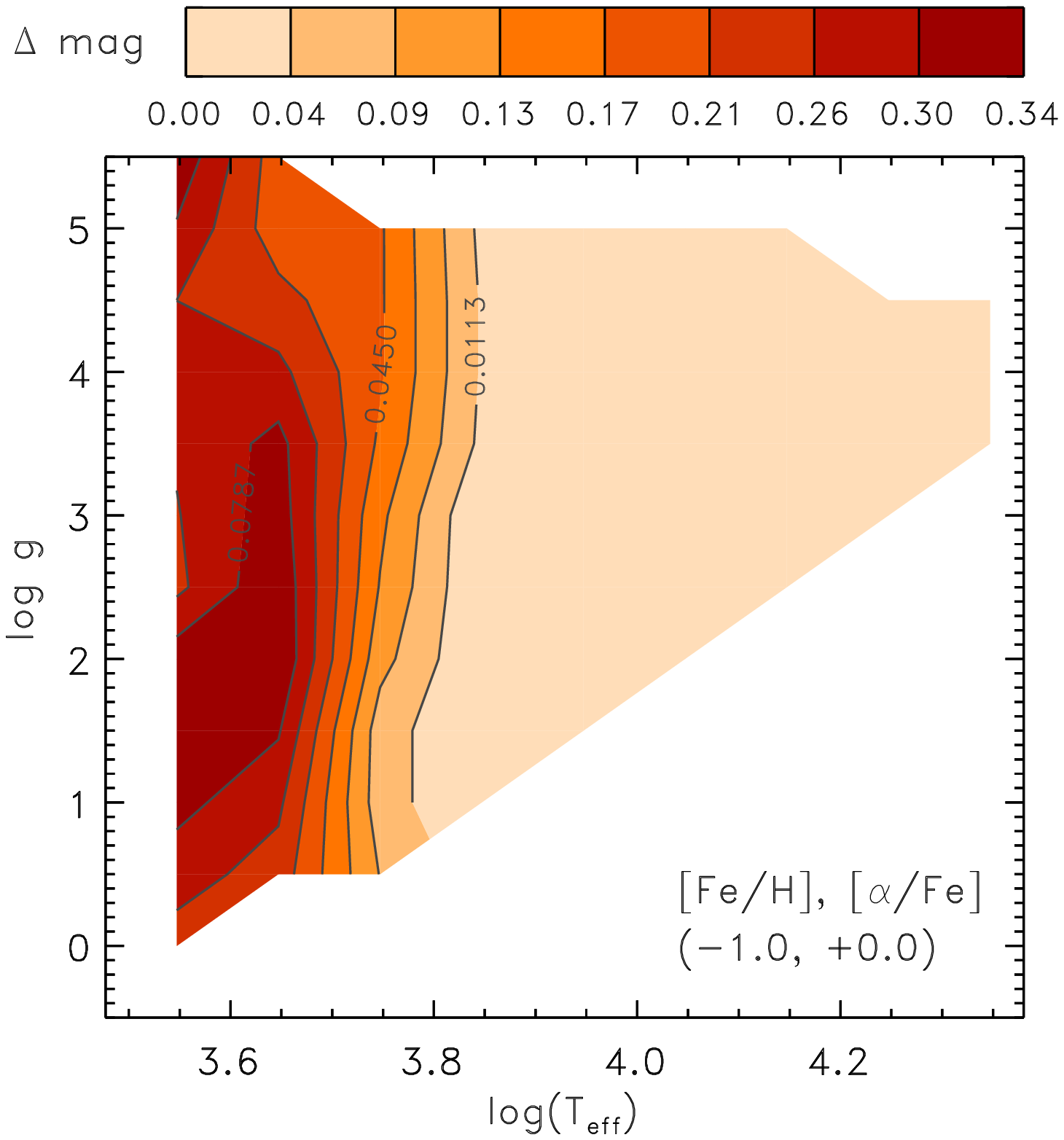}
\includegraphics[width=0.9\columnwidth]{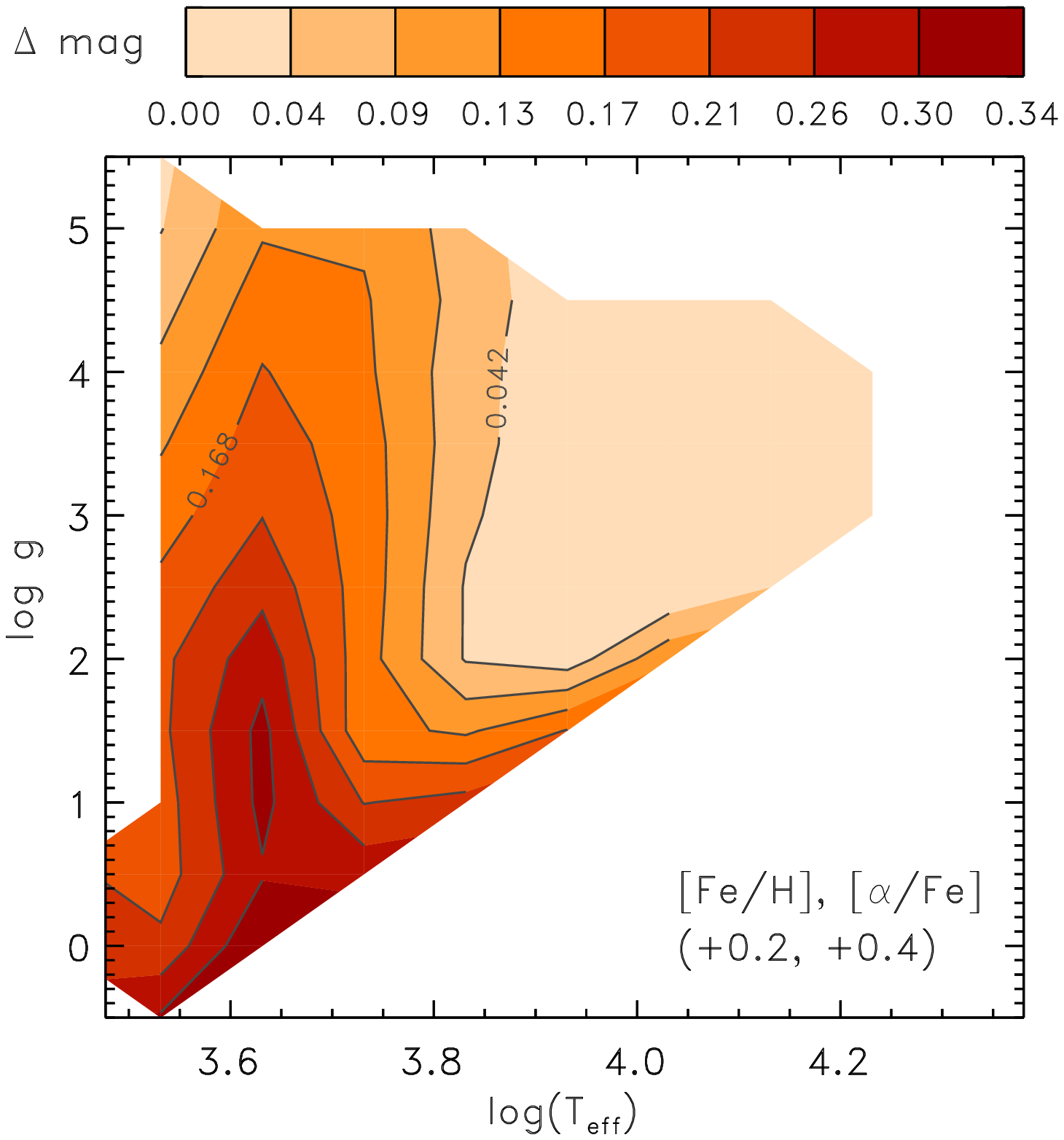}
\caption{Integrated flux differences between \mix{SED} and \mix{HIGHRES} models (see text in \S\ref{s:pl} for details), shown as contour maps in the plane log({\teff}) vs. {\logg}. Values are given in magnitudes. Blank areas correspond to parameters not covered by the present library. Maps are shown for mixtures \mix{m10p00} (top panel) and \mix{p02p04} (bottom panel).}
\label{fig_plmaps}
\end{center}
\end{figure}

A quantitative criteria was used to identify which stars are affected by the PLs in a non-negligible way: the fluxes in \mix{HIGHRES} and \mix{SED} models were integrated from 2500 to 6000\,{\AA} for the whole library. Fig. \ref{fig_plmaps} illustrates the difference in magnitudes between the \mix{HIGHRES} and \mix{SED} models in the plane log(\teff) vs. {\logg}, for the most metal poor and most metal rich mixture modelled in this work. The contour plots for the remaining mixtures are shown in the online Appendix A. 

\begin{figure}
\begin{center}
\includegraphics[width=\columnwidth]{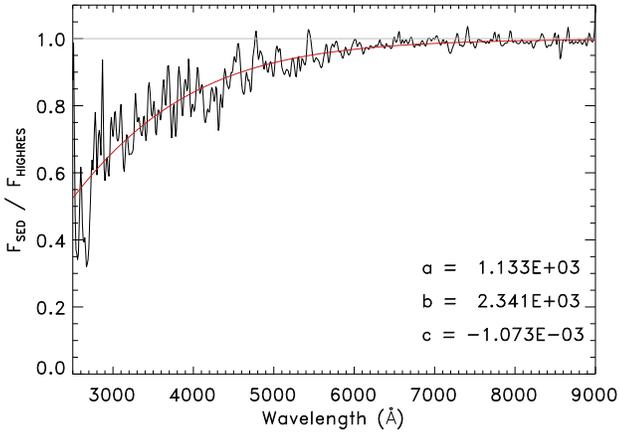}
\caption{Ratio between the \mix{SED} and the \mix{HIGHRES} models (black curve) for atmospheric parameters {\teff} = 5500\,K, {\logg} = +3.0 and mixture \mix{p00p00}. The best fit function is shown as red line, with the corresponding parameters shown in the panel (see Eq. \ref{eq1}). }
\label{fig_plcorrection}
\end{center}
\end{figure}

The stars with absolute differences between \mix{SED} and \mix{HIGHRES} models larger than or equal to 0.05\,mag were flagged. 
In those cases, flux ratios F$_{\rm ratio}$ = F$_{\rm SED}$ / F$_{\rm HIGHRES}$ were computed, after convolving both fluxes to a common spectral resolution of FWHM = 30\,{\AA} at 4000\,\AA. To each flux ratio, a function of the form:

\begin{center}
\begin{equation} \label{eq1}
y = \tanh[(x-a)/b]+c,
\end{equation}
\end{center}

\noindent was fitted, where $y$ is the flux ratio, $x$ is the wavelength in \AA, $a$, $b$ and $c$ are fitting constants. This functional form was chosen because it provides a smooth tracing of the continua ratio, being less sensitive to the residual line features than a polynomial or a spline function. A flux ratio and corresponding fit are illustrated in Fig. \ref{fig_plcorrection}. The IDL package MPFIT\footnote{\url{http://purl.com/net/mpfit}} \citep{MPFIT,MINPACK} was used for performing the fitting.  For the model stars with {\teff} $\le$ 4500\,K, where the fluxes in the blue end of the spectrum approach zero, masks with different weights were used to prevent the fitted functions to become negative. The regions with fainter fluxes due to molecular bands were given zero weight (2570 -- 2700, 3050 -- 3330, 4080 -- 4200 and 4940 -- 5160\,\AA), and two pseudo-continua regions at the blue end  were given a weight three times larger than the remaining intervals (2520-2550, 2830-2870\,\AA). 
 The full list of models flagged for continuum correction  and their corresponding  fitted coefficients are presented in the online Appendix B. A sample of the table is shown in Table \ref{t:pl_example}. It presents the atmospheric parameters (columns 1 to 3) and coefficients (columns 4 to 6) fitted to the ratios between \mix{SED} and \mix{HIGHRES} models (see Eq. \ref{eq1}).
 
\begin{table*}
\caption{\label{t:pl_example} Fitted coefficients to the ratios between \mix{SED} and \mix{HIGHRES} models. See Eq. \ref{eq1} in \S\ref{s:pl}. }
\begin{tabular}{ccccccc}
\hline
            &         &     &    & \multicolumn{3}{c}{Fitted coefficients} \\
{\teff} (K) & {\logg} (dex) & [Fe/H] & [$\alpha$/Fe] & a & b & c \\
\hline
3000 & +0.0 & -0.1 & +0.4  &  2.574E+03 & 1.166E+03 & 3.264E-03 \\
3000 & +0.0 & -0.3 & +0.4  &  2.595E+03 & 1.121E+03 & 7.698E-03 \\
3000 & +0.0 & -0.1 & +0.4  &  2.597E+03 & 1.132E+03 & 3.933E-02 \\
3000 & +0.0 & 0.0  &  0.0  &  2.574E+03 & 1.001E+03 & 5.873E-02 \\
3000 & +0.0 & 0.0  & +0.4 &  2.567E+03 & 1.146E+03 & -7.102E-03 \\
\hline
\end{tabular}\\
Full table in the on-line only manuscript\\
\end{table*}

\begin{figure}
\begin{center}
\includegraphics[width=\columnwidth]{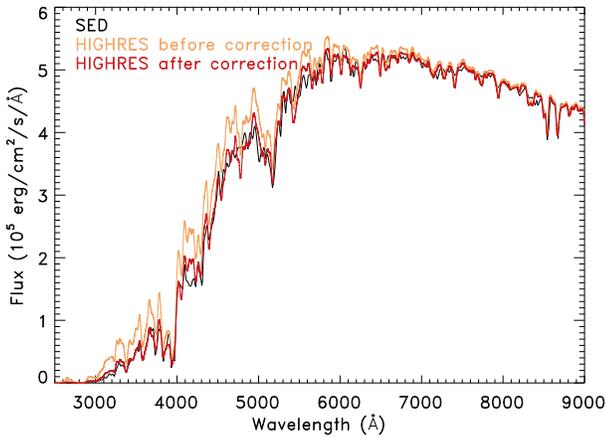}
\caption{Models are shown for a star with {\teff} = 4250\,K, {\logg} = 1.5\,dex and mixture \mix{m05p04}. Black curve shows the \mix{SED} model,  orange and red curves show \mix{HIGHRES} models before and after the correction for predicted lines, respectively. }
\label{fig:plcorrected}
\end{center}
\end{figure}

As a final step, the fitted functions were multiplied by the \mix{HIGHRES} models, resulting in models which kept the high spectral resolution features unhampered by the PLs, but flux distributions similar to the \mix{SED} models.
Fig. \ref{fig:plcorrected} illustrates the model spectra of a cool giant before and after the flux correction.

After these corrections were applied, colours in the Johnson-Morgan system measured on the \mix{HIGHRES} models reproduce the values measured on the \mix{SEDs} within the 0.02\,mag level. The median differences in colours predictions $\Delta$Colour = Colour$_{\rm SED}$ -- Colour$_{\rm HIGHRES}$ are shown in Table \ref{tab:plcor}, before and after the flux correction previously described.

\section{Comparisons between model predictions and observations}
\label{s:comparisons}

In this section the models are compared to observations in three distinct ways. The colour predictions of the \mix{SED} models are compared to a recent empirical calibration from bands U to K in \S\ref{s:colours}. The  \mix{HIGHRES} models are compared to the empirical library MILES \citep{MILES1,MILES2} in two domains: fluxes are compared in \S\ref{s:spectra} and atmospheric stellar parameters are compared in \S\ref{s:atmpars}.

\subsection{Broad band colours from \mix{SED} models}
\label{s:colours}

A convenient way of comparing the predicted \mix{SED} fluxes with observations is through broad-band colours.
In order to perform this comparison, representative pairs of {\teff} and {\logg} were chosen from two isochrones of a young and an old population with solar abundances (30\,Myr and 13\,Gyr). The isochrones are the same ones that stablished the coverage of the present stellar library, to be presented in our forthcoming stellar population models paper (Coelho et al. 2014, in prep.) 

The transformation to observed colours were done through the UBVRIJHK empirical calibration by \citet{worthey_lee11}\footnote{Colour-temperature table and interpolation program are available at \url{http://astro.wsu.edu/models/colorproj/colorpaper.html}}.  The authors adopted stars with accurately measured  photometry and known metallicity [Fe/H] to generate colour--colour relations that include the abundance dependence. Their data, taken from different sources in literature, were corrected for interstellar extinction and homogenised to a common system. 
A multivariate polynomial fitting program was applied to the data, and the final results are colour-temperature relations as a function of gravity and abundance.

  \begin{table}
  \centering
  \caption{\label{tab:plcor}Median differences in colour predictions between \mix{SED} and \mix{HIGHRES} models, for stars flagged as being affected by PL blanketing.}
  \begin{tabular}{@{} ccc @{}}
    \hline
                     & \multicolumn{2}{c}{$\Delta$ Colour (\mix{SED}-\mix{HIGHRES})}\\
    Colour & Before correction & After correction  \\ 
    \hline
U--B &     0.171 &    -0.018 \\
B--V &     0.084 &     0.025 \\
V--R &     0.026 &     0.007 \\
  \hline
  \end{tabular}
  \end{table}

The magnitudes predicted by \mix{SED} models were measured using the task \texttt{SBANDS} in IRAF\footnote{IRAF is distributed by the National Optical Astronomy Observatories,
    which are operated by the Association of Universities for Research
    in Astronomy, Inc., under cooperative agreement with the National
    Science Foundation. \url{http://iraf.noao.edu/}} \citep{IRAF1,IRAF2}, adopting the filter transmission curves of the photometric systems adopted in \citet{worthey_lee11}. Zero-point corrections were applied to the model magnitudes using the Vega model by \citet{castelli_kurucz94}\footnote{\url{http://wwwuser.oat.ts.astro.it/castelli/vega.html}}, resampled to the wavelength sampling of the \mix{SED} models. 
Vega magnitudes were adopted to be (G. Worthey, priv. comm.):
U$_{\rm Johnson}$ = 0.02, B$_{\rm Johnson}$ = 0.03, V$_{\rm Johnson}$ = 0.03, R$_{\rm Cousin}$ = 0.039, I$_{\rm Cousin}$ = 0.035, J$_{\rm Bessell}$ = 0.02, H$_{\rm Bessell}$ = 0.02, K$_{\rm Bessell}$ = 0.02.

\begin{figure*}
\begin{center}
\includegraphics[width=1.5\columnwidth]{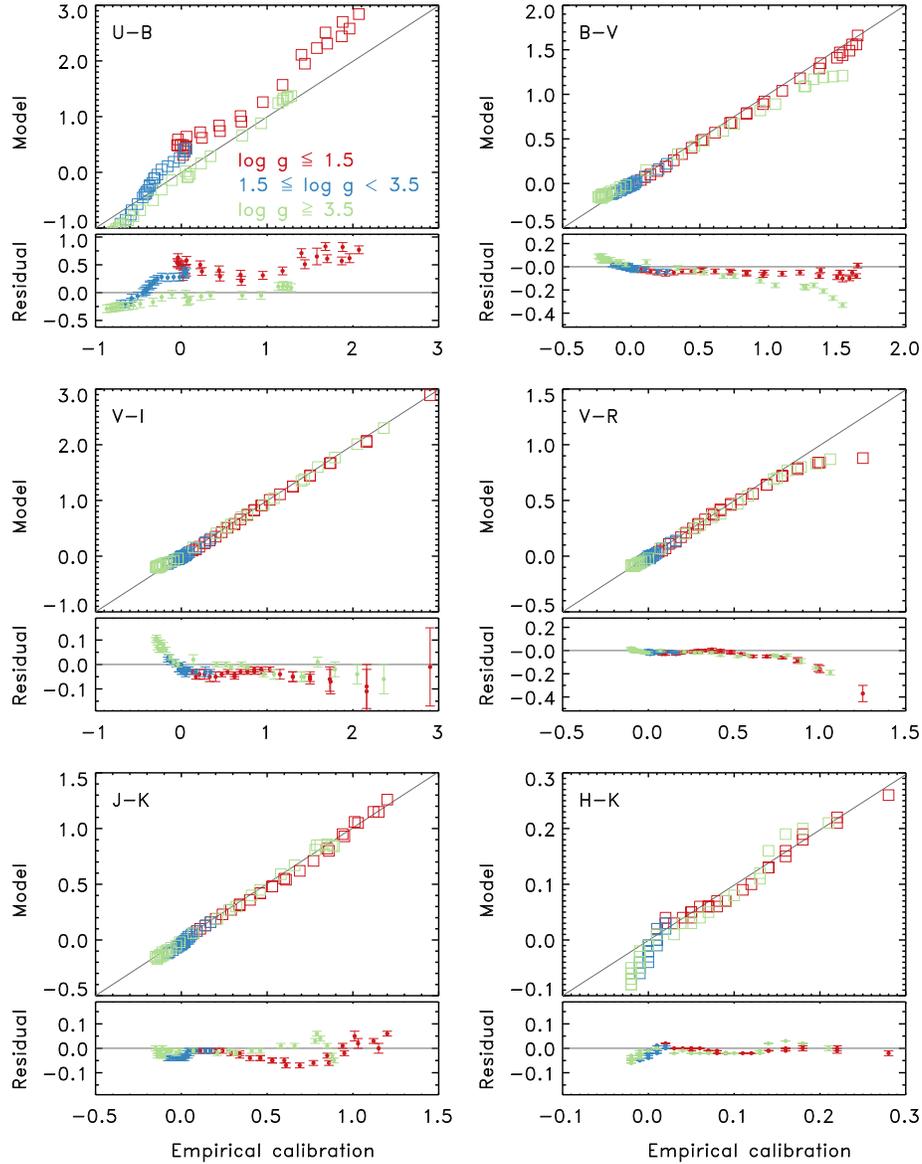}
\caption{Comparison between colours predicted by \mix{SED} models (\emph{y}-axis) and colours from the empirical relations derived by \citet{worthey_lee11} (\emph{x}-axis). Residuals are shown below each panel, where the error bars are the uncertainties of the empirical relations. The  colours illustrate different intervals in {\logg}, as indicated in the upper-left panel. Data correspond to solar abundances (mixture \mix{p00p00}), and {\teff}--{\logg} pairs chosen from an isochrone of 30\,Myr and an isochrone of 14\,Gyr.  }
\label{fig_colours1}
\end{center}
\end{figure*}

Comparisons between the empirical relation and the model predictions are given in Fig. \ref{fig_colours1}.
The coloured symbols indicate the \mix{SED} predictions in different {\logg} intervals, as indicated in the figure. 
Residuals (model minus empirical) are shown below each panel, where the error bars indicate the uncertainties of the \citet{worthey_lee11} calibration.
The behaviour shown by the \mix{SED} models is very similar to what was obtained by \citet{MC07} for \citet{ATLASODFNEW} models, as expected given that the procedure and ingredients of both set of models are the same (the only differences for the solar mixture set of models is the molecular line list for H$_2$O, corrected in Feb/2012 by R.  Kurucz).
Table \ref{t:difcolors} shows the average differences between model and empirical relations.

  \begin{table}
  \centering
  \caption{\label{t:difcolors} Average differences in colour between \mix{SED} prediction and the empirical calibration by \citet{worthey_lee11}.}
  \begin{tabular}{@{} cc @{}}
    \hline
    Broad-band colour & $\Delta$ Colour (Model -- Empirical) \\ 
    \hline
U--B &   0.168 \\
B--V &  -0.032 \\ 
V--I &  -0.011 \\
V--R &  -0.031 \\
J--K &  -0.018 \\
H--K &  -0.015 \\
  \hline
  \end{tabular}
  \end{table}

The \mix{SED} predictions reproduce the empirical calibration for a large fraction of the colour ranges. Exceptions are {\logg} $\le$ 3.5 for U--B colour, blue extremes of the B--V and V--I panels and red extremes of the B--V (dwarfs) and V--R panels.
To isolate the reasons for the noted discrepancies between model predictions and empirical calibration is beyond the purpose of the present paper, but several detailed discussions existing in literature apply to the current models \citep[e.g.][]{bessell+98,kucinskas+05,MC07,plez11proc}.  
\citet{kucinskas+05}, for example, present synthetic broad-band photometric colours for late-type giants based on synthetic spectra calculated with the PHOENIX code \citep{PHOENIX05}, and carefully explored the effect of several ingredients and assumptions (such as molecular opacities, gravity, micro-turbulent velocity, and stellar mass) on the resulting model colours. They also compared PHOENIX predictions with ATLAS9 \citep{ATLASODFNEW} and MARCS \citep{MARCS08} models. Their work confirms that synthetic colours of PHOENIX, MARCS, ATLAS9 agree to within $\Delta${\teff} $\sim$ 100\,K over a large range of effective temperatures, despite the fact that PHOENIX models assume spherical geometry while ATLAS9 colours are obtained from plane-parallel model atmospheres. Nevertheless, they noted that convection may influence photometric colours in a non-negligible way. The difference between synthetic colours calculated with a fully time-dependent 3D hydrodynamical model atmosphere and those obtained with the conventional 1D model may reach up to several tenths of  magnitude in certain photometric colours (e.g., $\Delta$(V -- K) $\sim$ 0.2\,mag). Also, the authors showed that the B--V colour is the more complex of all colours investigated (they did not study the U--B colour): while the agreement between observed and synthetic colours is good at higher effective temperatures, all temperature-colours scales tend to disagree below $\sim$3800\,K.


\subsection{Fluxes from \mix{HIGHRES} models}
\label{s:spectra}

In this section the \mix{HIGHRES} models are compared to observed spectra from the empirical stellar library MILES \citep{MILES1,MILES2}.
The reasons for choosing MILES as the proxy empirical library among the multitude of available libraries\footnote{See the list maintained by David Montes at \url{http://www.ucm.es/info/Astrof/invest/actividad/spectra.html}} are the following: 

\begin{itemize}
\item MILES is currently the standard empirical library for use in stellar population models \citep[][Charlot \& Bruzual, in prep.]{vazdekis+10, martin-hernandez+10proc, maraston_stromback11};
\item it has an optimal coverage of the HR diagram with $\sim$ 1000 stars; currently its coverage is only  rivalled by ELODIE library \citep[][and references within]{ELODIE3}, which nevertheless has a shorter wavelength range and a poorer coverage of giants stars, which dominate over dwarfs in the integrated light of populations, and;
\item the [Fe/H] vs [$\alpha$/Fe] relation for MILES stars was well characterised in \citet{milone+11}, making MILES highly suitable to be compared with the library of the present work.
\end{itemize}

The correspondence between models and observations is naturally done via the atmospheric parameters. Accurate atmospheric parameters are also a key aspect to link the stellar spectral library to stellar evolution prescriptions, another crucial ingredient 
of a stellar population model. 
For instance, \citet{percival_salaris09} performed an interesting investigation of the possible impact of systematic uncertainties in atmospheric parameters on integrated spectra of stellar populations. Those authors raised a caution by showing that small systematic differences between the atmospheric parameters scales can mimic non-solar abundance ratios or multi-populations in the analysis of integrated spectra.
With the goal of performing statistical comparisons between model and empirical spectra, 
 an effort was made to estimate realistic uncertainty intervals for the atmospheric parameters adopted in the empirical library. 

\subsubsection{Uncertainties on atmospheric parameters}

  \begin{table}
  \centering
  \caption{\label{tab:error}Errors in the atmospheric parameters adopted for the comparison between model and observed spectra.}
  \begin{tabular}{@{} rcc @{}}
    \hline
                     & \multicolumn{2}{c}{Adopted errors}\\
    {\teff} interval & $\sigma$(\teff) & $\sigma$(\logg)  \\ 
    \hline
    3000 -- 4000   &   120   &    0.3\\ 
    4000 -- 5000   &   120   &    0.2\\  
    5000 -- 7000   &   120   &    0.1\\  
    7000 -- 9000   &   250   &    0.1\\  
    9000 -- 10000  &   250   &    0.2\\  
   10000 -- 13000  &   400   &    0.2\\  
   13000 -- 16000  &   650   &    0.2\\  
   16000 -- 18000  &   850   &    0.2\\  
   18000 -- 21000  &  1000   &    0.2\\  
   21000 -- 23000  &  1400   &    0.2\\  
   above 23000     &  3000   &    0.2\\  
  \hline
  \end{tabular}
  \end{table}
  
Often the parameters of observed stellar spectra 
are derived by comparison to models, or to calibrations which are largely based on 
models \citep[e.g.][]{bessell+98}. On the other hand, modellers of stellar spectra need stars with \teff~and 
\logg~derived by fundamental ways \citep[independent or weakly dependent on models, see 
e.g.][]{cayrel02} in order to test and calibrate the models. In the case of temperatures, for 
example, direct estimation of \teff~is possible for close stars if the angular diameter of a 
star is known (interferometric measurements or lunar occultations; e.g. \citealt[][]{code+76,dibenedetto93,kervella+04,vanbelle+09}). Recent 
determinations are able to determine \teff~with a typical accuracy of 5\% (see e.g. compilations in \citealt{jerzykiewicz_zakowicz00,torres+10}). Moreover, many M giants ({\teff} $\apprle$ 4000\,K) are long period variables \citep[e.g.][]{banyai+13}, and one may wonder if the published values for atmospheric parameters correspond to the epoch of observation in the empirical library. \citet{kucinskas+05} pointed out that, in their 
search for published interferometric effective temperatures of late-type giants in the solar neighbourhood, none non-variable giant with effective temperature lower than {\teff} $\sim$ 3400\,K was found.

In the case of empirical libraries such as MILES, methods of deriving the atmospheric parameters 
based on a reference sample of well studied stars \citep[e.g.][]{katz+98,soubiran+98}  
guarantee homogeneous estimations, and were indeed adopted by e.g. \citet{ELODIE,MILES2}.
Homogeneous estimations do 
not guarantee, however, against systematic errors, if the parameters of the reference 
stars are affected by undetected systematic deviations. Moreover, the reference stars 
usually encompass a limited range of spectral types, and outside this range the derived 
parameters are less reliable.

Atmospheric parameters for the MILES stars were first compiled by
\citet{MILES2}, but they did not provide star-by-star errors. More recently, \citet{prugniel+11} re-derived the parameters and provided fitting errors for the majority of stars. These errors correspond to the internal precision of the method adopted and might not give a fair assessment of the accuracy of the parameters. From a different perspective, a recent compilation of atmospheric parameters derived from fundamental methods is given in \citet{torres+10}, where uncertainties in {\teff} typically range from 2 to 5\%.

\begin{figure}
\begin{center}
\includegraphics[width=1.0\columnwidth]{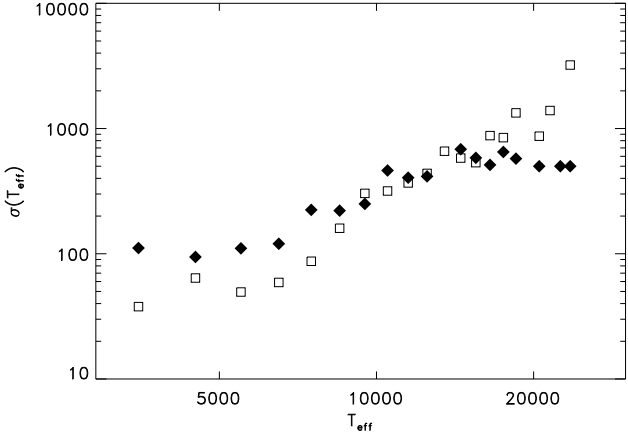}
\caption{Relation between typical errors in {\teff} 
as a function of {\teff} from two sources in literature: errors from \citet{prugniel+11} are shown as open squares and errors from \citet{torres+10} are shown as filled diamonds. Points correspond to average errors in intervals of {\teff} of 1000\,K.}
\label{fig_errorsteff}
\end{center}
\end{figure}

In order to compare uncertainties quoted in both works, for every interval  of 1000\,K in {\teff}, the average error from \citet{torres+10} and from \citet{prugniel+11} were computed (only stars around solar metallicity were considered -0.15 $\le$ [Fe/H] $\le$ 0.15). Results are illustrated in Fig. \ref{fig_errorsteff}, and three regimes are seen: 
$(a)$ below {\teff} $\sim$ 11000\,K, Prugniel et al. uncertainties are smaller than in Torres et al; 
$(b)$ between $\sim$ 12000 and 17000\,K, the uncertainties from both work are comparable, and;
$(c)$ above {\teff} $\sim$ 18000\,K uncertainties from Prugniel et al. are larger than in Torres et al.
This trend likely reflects the fact that in MILES, hotter stars are relatively sparse and F, G and K are more abundant, allowing the fitting in this latter regime to be more precise. Nevertheless, it is unlikely that the final error (considering both precision and accuracy) is smaller than the uncertainty obtained by determinations from fundamental methods such as the ones in Torres et al. 

Through the remaining of this work, the error in temperature $\sigma$({\teff}) per {\teff} interval was assumed to be the average errors quoted by Prugniel et al., except in the first regime noted in Fig. \ref{fig_errorsteff}, where average errors from Torres et al. were adopted. For the case of errors in {\logg}, average errors from Prugniel et al. were adopted for the whole range of parameters, as they are typically an order of magnitude larger than the errors from fundamental methods. The error in [Fe/H] was conservatively adopted to be 0.15\,dex \citep[e.g.][]{soubiran+98}. The final uncertainties adopted per {\teff} interval are given in 
Table \ref{tab:error}.

\subsubsection{Flux comparisons}

For each pair [{\teff}, {\logg}] existing in mixture \mix{p00p00} (solar metalicity) in the model library, MILES library (adopting parameters from \citealt{prugniel+11}) was searched for stars with parameters within intervals given by the uncertainties in Table \ref{tab:error}. 
To compare each model spectrum to several empirical spectra inside the uncertainty intervals serve two purposes: to take into account how uncertainties in the atmospheric parameters affect the fluxes, and to smooth out chemical peculiarities from individual stars. Adopting some empirical stars inside parameters uncertainties helps establishing confidence limits for evaluating the quality of the model.

Relatively few intervals were found containing at least three empirical stellar spectra. 
 Above {\teff} = 12000\,K, at most two stellar spectra are found in a given interval. These intervals, in total 37, are reported in Table \ref{t:milespar}. The table lists the {\teff} and {\logg} intervals studied (columns 1 and 2), the number of MILES stars within each interval (column 3), the corresponding average  {\teff} and {\logg} from MILES stars (columns 4 and 5), and the mean  
 absolute deviation $\overline\Delta$ between the model spectrum and the averaged empirical spectrum (column 6).
$\overline\Delta$ is defined as: 

\begin{equation}
\overline\Delta = \frac{1}{N_{\lambda}} \sum_{\lambda}{\Big | \frac{[f_{\rm model}(\lambda)-f_{\rm obs}(\lambda)]}{f_{\rm obs(\lambda)}}\Big |}
\label{eq2}
\end{equation}

\noindent where 
\textit{N$_\lambda$} is the number of pixels in each spectrum, $f_{\rm model}$ is the model cpectrum and $f_{\rm obs}$ is the average empirical spectrum. Before computing the $\overline\Delta$, model and empirical spectra were brought to a common wavelength and flux scale: the model spectra were convolved to MILES spectral resolution \citep{fbarroso+11}, model and empirical spectra were resampled to a common wavelength sampling of 0.5\,{\AA}, and each spectrum was normalised to $\int F_\lambda {\rm d}\lambda = 1$.

\begin{table*}
  \centering
    \caption{\label{t:milespar}For a given interval in atmospheric parameters (defined in columns 1 and 2), the number of stars in MILES inside that interval is given in column 3. Columns 4 and 5 indicate, respectively, the average {\teff} and {\logg} of the MILES stars. Column 6 shows the average absolute deviation (equation \ref{eq2}) between the model spectrum and the observed stars. Only stars with iron abundances close to solar are considered (-0.15 $\le$[Fe/H] $\le$ 0.15). Only intervals where at least 3 MILES stars exist are shown, with the exception of stars hotter than 12000\,K, where at maximum 2 stars exist per interval. These intervals were used in the comparisons presented in Figs. \ref{fig_miles}, \ref{fig:residuals} and figures in the Apendix (electronic edition of this manuscript only). }
  \begin{tabular}{@{} cccccc @{}}
    \hline
    {\teff} interval (K)& {\logg} interval (dex)& \# of stars in MILES & Average {\teff} & Average \logg & $\overline\Delta$  \\ 
    \hline
 3200 $\pm$ 120 &  0.50 $\pm$ 0.30           &  5 &  3244 &  0.5  & 54.8\% \\
 3400 $\pm$ 120 &  0.50 $\pm$ 0.30           &  8 &  3400 &  0.7  & 30.0\% \\
 3400 $\pm$ 120 &  1.00 $\pm$ 0.30           &  5 &  3441 &  0.8  & 34.6\% \\
 3600 $\pm$ 120 &  1.00 $\pm$ 0.30           &  6 &  3629 &  1.0  & 18.4\% \\
 3800 $\pm$ 120 &  1.00 $\pm$ 0.30           &  10 & 3798 &  1.1  & 13.1\% \\
 3800 $\pm$ 120 &  1.50 $\pm$ 0.30           &  9  & 3830 &  1.4  & 11.9\% \\
 4000 $\pm$ 120 &  1.00 $\pm$ 0.20           &  5  & 3973 &  1.0  & 10.9\% \\
 4000 $\pm$ 120 &  1.50 $\pm$ 0.20           &  8  & 4011 &  1.6  &  9.2\% \\
 4000 $\pm$ 120 &  2.00 $\pm$ 0.20           &  4  & 3983 &  2.0  &  9.0\% \\
 4250 $\pm$ 120 &  1.50 $\pm$ 0.20           &  4  & 4232 &  1.6  &  8.6\% \\
 4250 $\pm$ 120 &  2.00 $\pm$ 0.20           &  6  & 4252 &  2.0  &  8.7\% \\
 4250 $\pm$ 120 &  4.50 $\pm$ 0.20           &  5  & 4288 &  4.5  &  8.8\% \\
 4500 $\pm$ 120 &  2.50 $\pm$ 0.20           &  8  & 4559 &  2.5  &  6.7\% \\
 4500 $\pm$ 120 &  4.50 $\pm$ 0.20           &  3  & 4449 &  4.6  &  7.1\% \\
 4750 $\pm$ 120 &  2.50 $\pm$ 0.20           &  15 & 4767 &  2.6  &  5.2\% \\
 4750 $\pm$ 120 &  4.50 $\pm$ 0.20           &  4  & 4738 &  4.6  &  5.4\% \\
 5000 $\pm$ 120 &  2.50 $\pm$ 0.10           &  3  & 4914 &  2.5  &  4.2\% \\
 5250 $\pm$ 120 &  4.50 $\pm$ 0.10           &  7  & 5256 &  4.5  &  3.9\% \\
 5500 $\pm$ 120 &  4.50 $\pm$ 0.10           &  4  & 5442 &  4.5  &  2.6\% \\
 6000 $\pm$ 120 &  4.00 $\pm$ 0.10           &  5  & 6054 &  4.0  &  3.8\% \\
 6250 $\pm$ 120 &  4.00 $\pm$ 0.10           &  7  & 6248 &  4.0  &  2.1\% \\
 6500 $\pm$ 120 &  4.00 $\pm$ 0.10           &  10 & 6499 &  4.1  &  1.9\% \\
 6750 $\pm$ 120 &  4.00 $\pm$ 0.10           &  6  & 6726 &  4.0  &  1.6\% \\
 7000 $\pm$ 250 &  4.00 $\pm$ 0.10           &  9  & 7000 &  4.0  &  1.4\% \\
 7250 $\pm$ 250 &  4.00 $\pm$ 0.10           &  9  & 7245 &  4.0  &  1.7\% \\
 7500 $\pm$ 250 &  4.00 $\pm$ 0.10           &  6  & 7387 &  4.0  &  1.4\% \\
 7750 $\pm$ 250 &  4.00 $\pm$ 0.10           &  3  & 7777 &  4.0  &  1.5\% \\
 8000 $\pm$ 250 &  4.00 $\pm$ 0.10           &  3  & 7969 &  3.9  &  1.6\% \\
 9750 $\pm$ 250 &  4.00 $\pm$ 0.20           &  3  & 9640 &  3.9  &  2.1\% \\
10000 $\pm$ 400 &  4.00 $\pm$ 0.20           &  3 & 10080 &  3.9  &  3.3\% \\
10500 $\pm$ 400 &  4.00 $\pm$ 0.20           &  3 & 10388 &  3.9  &  2.3\% \\
10750 $\pm$ 400 &  4.00 $\pm$ 0.20           &  6 & 10978 &  3.9  &  2.0\% \\
11000 $\pm$ 400 &  4.00 $\pm$ 0.20           &  8 & 11052 &  3.9  &  1.3\% \\
11250 $\pm$ 400 &  4.00 $\pm$ 0.20           &  8 & 11144 &  4.0  &  1.1\% \\
11500 $\pm$ 400 &  4.00 $\pm$ 0.20           &  3 & 11337 &  4.0  &  2.1\% \\
18000 $\pm$1000 &  3.50 $\pm$ 0.20           &  2 & 18170 &  3.7  &  2.4\% \\
21000 $\pm$1400 &  4.00 $\pm$ 0.20           &  2 & 20873 &  3.8  &  3.0\% \\
  \hline
  \end{tabular}
\end{table*}

\begin{figure*}
\begin{center}
\includegraphics[bb=0 150 588 758,width=1.9\columnwidth]{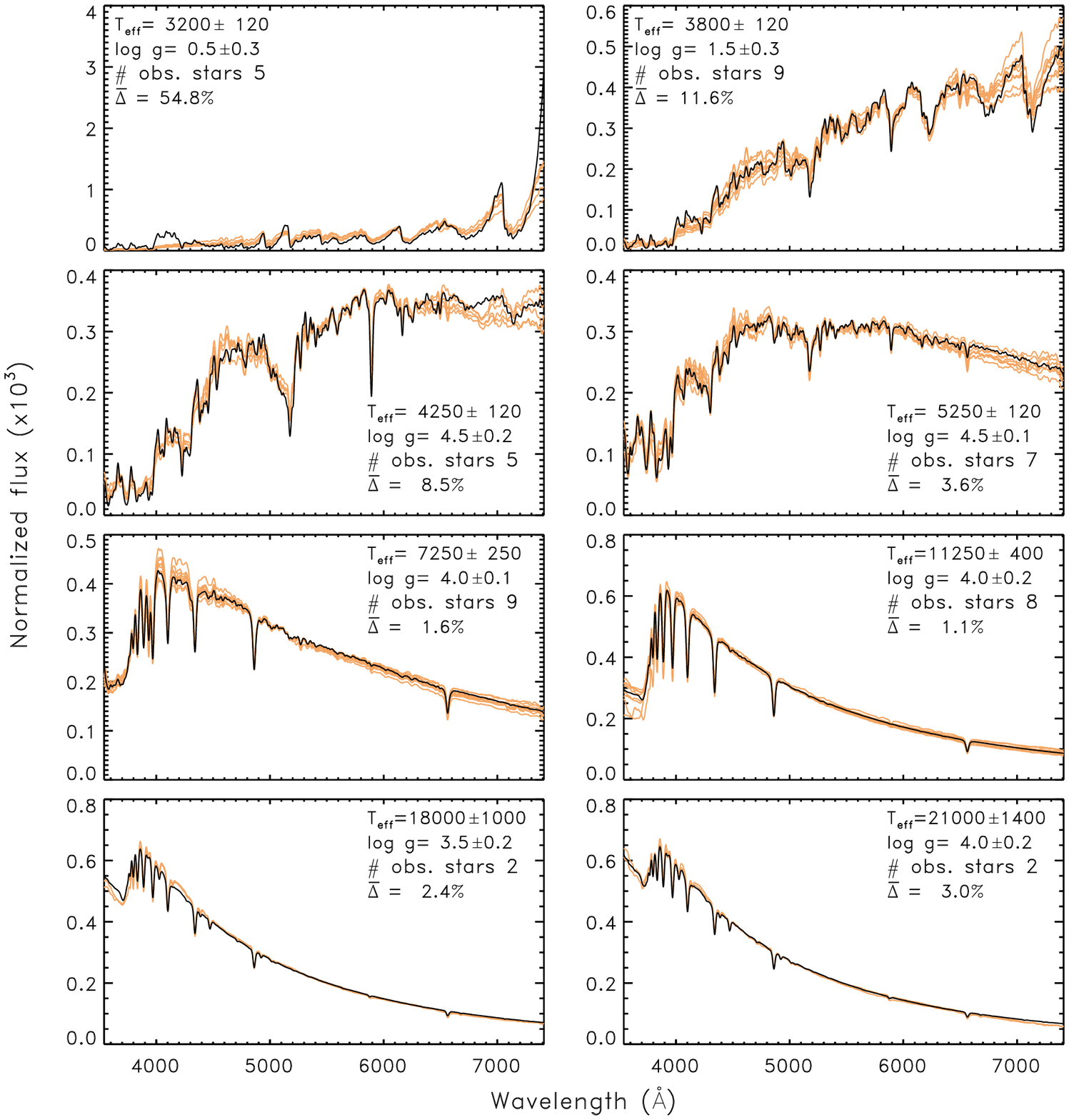}
\caption{\mix{HIGHRES} models (black lines) are shown along observed spectra from MILES (colour lines). Each stellar flux was normalised to $\int F_\lambda {\rm d}\lambda = 1$ and multiplied by 10$^3$. Intervals of atmospheric parameters are indicated in each panel. For each interval, all solar-metallicity MILES stars inside the indicated parameters ranges are shown. All spectra were broadened to low resolution for better visualisation. Comparisons for other parameter intervals are shown in the Apendix A (online manuscript).}
\label{fig_miles}
\end{center}
\end{figure*}

Comparisons between model and empirical spectra for eight of these intervals are shown in Fig. \ref{fig_miles}.
In this figure, the black curves represent the model \mix{HIGHRES} spectra and the coloured curves show the empirical spectra within a given interval of parameters (as indicated in the panels). These panels were chosen to span the whole set of temperatures and illustrate pairs [{\teff}, {\logg}] with at least 5 empirical spectra. Exception was made for the last two panels (hottest temperatures), where no such interval exists. The coolest solar metalicity star in MILES, adopting \citet{prugniel+11} parameters, has {\teff} $\sim$ 3200\,K. The comparisons for other intervals reported in Table \ref{t:milespar} are shown in Appendix A (online manuscript). 

From the $\overline\Delta$ values computed, it is seen that model stars with {\teff} $\ge$ 4750 reproduce observed fluxes within $\sim$ 5\%. The model spectra systematically deviates from empirical fluxes as {\teff} drops below this limit, reaching $\sim$ 50\% at the coolest interval. Stars with {\teff} $\ge$ 6250 are typically reproduced within 2\%. 

\begin{figure*}
\begin{center}
\includegraphics[width=1.9\columnwidth]{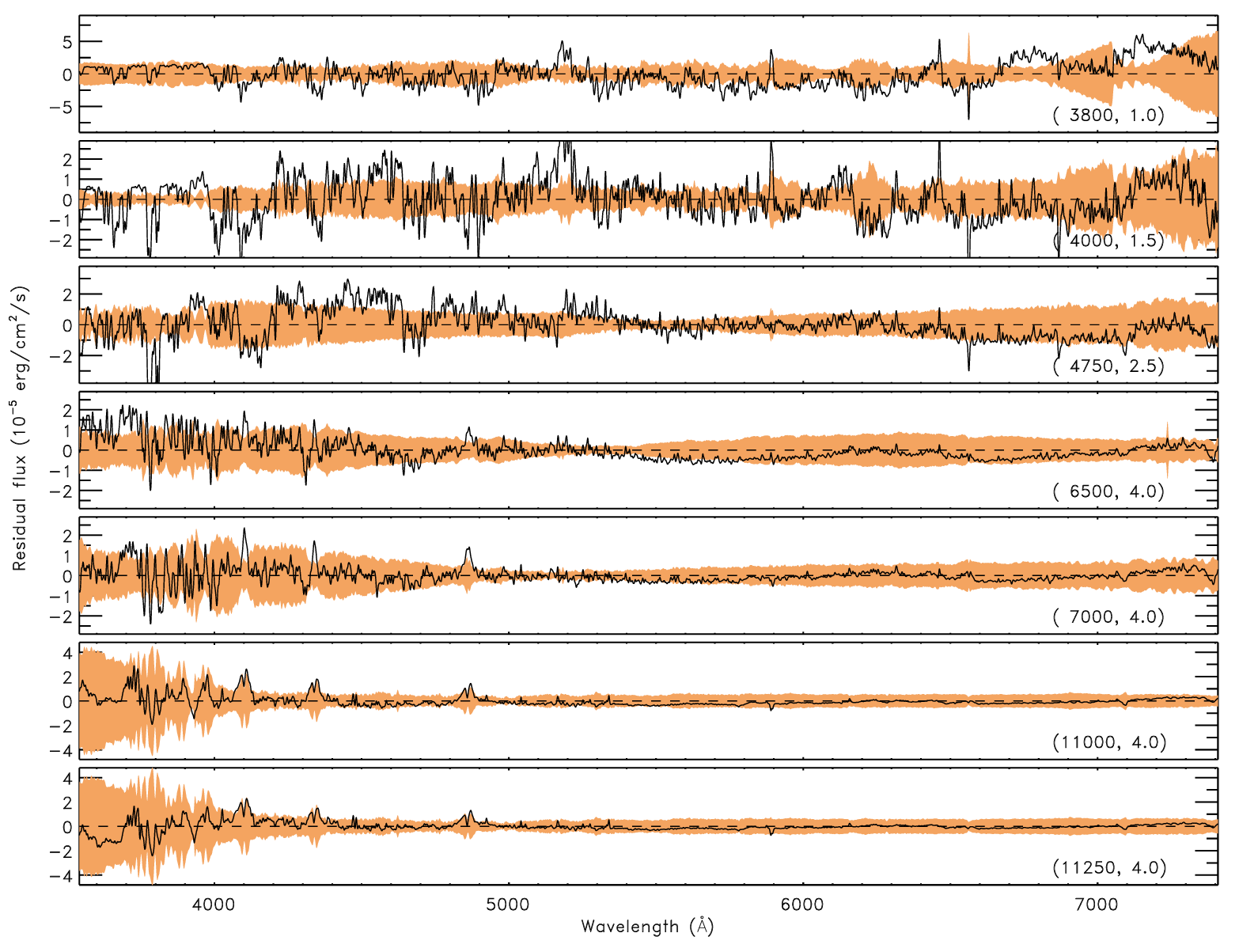}
\caption{Black curves show the flux difference between the average MILES star in a given interval of atmospheric parameters (see Table \ref{t:milespar}) minus the correspondent \mix{HIGHRES} model (atmospheric parameters indicated in each panel).  The coloured area in each panel indicate the r.m.s. of the observed spectra. }
\label{fig:residuals}
\end{center}
\end{figure*}

For the intervals with at least 8 MILES stars within, a root mean squared r.m.s. observed spectrum was computed. The difference between averaged observed spectrum and model spectrum is shown in Fig. \ref{fig:residuals} as black curves, for seven intervals of atmospheric parameters. The coloured areas correspond to $\pm$ r.m.s limits, derived from the observed stars. 
At first approximation, residuals below the r.m.s. area correspond to missing opacities in the model, while residuals above the r.m.s. area correspond to lines excessively strong. Some prominent regions, seen in more than one interval, are: 
\begin{enumerate}
\item features below 4200\,{\AA} corresponding to missing opacities, noted in {\teff} up to 4750\,K;
\item evidence for excessive opacity near 4300, 4700 and 5200\,{\AA} (seen in particular in {\teff} = 4000 and 4750\,K), potentially related to bands of CH, C$_2$ and MgH respectively, and;
\item too strong core of H lines, in {\teff} = 6500\,K and above, potentially related to the fact that core of very strong lines are formed in N-LTE in the chromosphere layers. 
\end{enumerate}

In the case of item (ii) above, it is important to note that the effect was seen in the comparisons with cool giants only (there were no intervals with at least 8 spectra of cool dwarfs). It would be interesting to further investigate if the effects at CH A--X and C$_2$ bands could be related to other effects such as non-solar abundances of C and N due to dredge up. Different treatments of convection in the model atmosphere may also affect the intensity of molecular features \citep[e.g.][M. P. Diaz, priv. comm.]{kucinskas+05}.

\subsection{Model versus fitted atmospheric parameters}
\label{s:atmpars}

An alternative way to compare model and observations is in the space of the atmospheric parameters. \citet{bertone+08} compared the high resolution spectrum of the Sun to a small grid of theoretical 
libraries, and derived the solar parameters using the theoretical grid as reference stars. 
The parameters derived for the Sun had offsets with respect to the real values of 
$\Delta$\teff~= +80K, $\Delta$\logg~= +0.5 and $\Delta$[Fe/H]= -0.3. These offsets quantify 
the accuracy of the theoretical library in a scale that can be directly compared to the 
uncertainties of the atmospheric parameters in empirical libraries. 

Inspired by Bertone et al. results, a similar exercise is done in this work, inverting the role of the model and observed spectra: atmospheric parameters for the model spectra were obtained using as template reference the MILES stars and their derived atmospheric parameters. This is a convenient way of performing this exercise given the deployment of a spectral interpolator based on MILES stars \citep{prugniel+11}, to be used with the public code \textsc{ULySS} \citep{ULYSS}. \textsc{ULySS} is a software package which started as an adaptation from pPXF code by \citet{pPXF}, and performs spectral fitting in two astrophysical contexts: the determination of stellar atmospheric parameters and the study of the star formation and chemical enrichment history of galaxies. In ULySS, an observed spectrum is fitted by a model (expressed as a linear combination of components) through a non-linear least-squares minimisation. For the present study, the model is the MILES interpolator by \citet{prugniel+11}.

 \begin{table}
  \centering
    \caption{[Fe/H] vs. [$\alpha$/Fe] relation for MILES library, obtained from the results by \citet{milone+11}. The first column shows the [Fe/H] values available in the model library. The average [$\alpha$/Fe] in the second column was computed from the stars in MILES with [Fe/H] inside $\pm$ 0.15\,dex of the model values, weighted by the inverse of the measurement error.  }
  \begin{tabular}{S[table-format=3.2] S[table-format=3.2]}
    \hline
    \multicolumn{1}{c}{[Fe/H]} & \multicolumn{1}{c}{[$\alpha$/Fe]}  \\ 
    \hline
     -1.3  &   0.36 \\
     -1.0  &   0.42\\
    -0.8   &  0.38\\
    -0.5   &  0.21\\
    -0.3   &  0.12\\
    -0.1   & 0.05\\
      0.0  &  0.02\\
     0.2   & 0.005\\
  \hline
  \end{tabular}
  \label{tab:milone}
\end{table}

Before the comparisons were performed, a sample selection was needed to evaluate which of the \mix{HIGHRES} mixtures were suitable to be compared to MILES library. Being empirical, MILES is biased to the [Fe/H] vs [$\alpha$/Fe] relation of the solar neighbourhood, while [Fe/H] and [$\alpha$/Fe] in the model library are varied independently. 
Recently, \citet{milone+11} obtained measurements of [Mg/Fe] for 76\% of MILES stars via compilation of values derived in literature and their own spectroscopic analysis. From their results, the mean values of [$\alpha$/Fe] were computed for each value of [Fe/H] available in the present theoretical grid. Intervals of $\pm$\,0.15\,dex were allowed in [Fe/H] and the average [$\alpha$/Fe] values were weighted by the inverse of the  quoted errors. Results are shown in Table \ref{tab:milone}. From those values, and assuming that at first approximation [$\alpha$/Fe] = [Mg/Fe], the mixtures of the present library which can be safely compared to MILES are: \mix{m13p04}, \mix{m10p04}, \mix{m08p04}, \mix{p00p00} and \mix{p02p00} (see Table \ref{tab_mixes}).

\begin{figure*}
\begin{center}
\includegraphics[width=0.9\columnwidth]{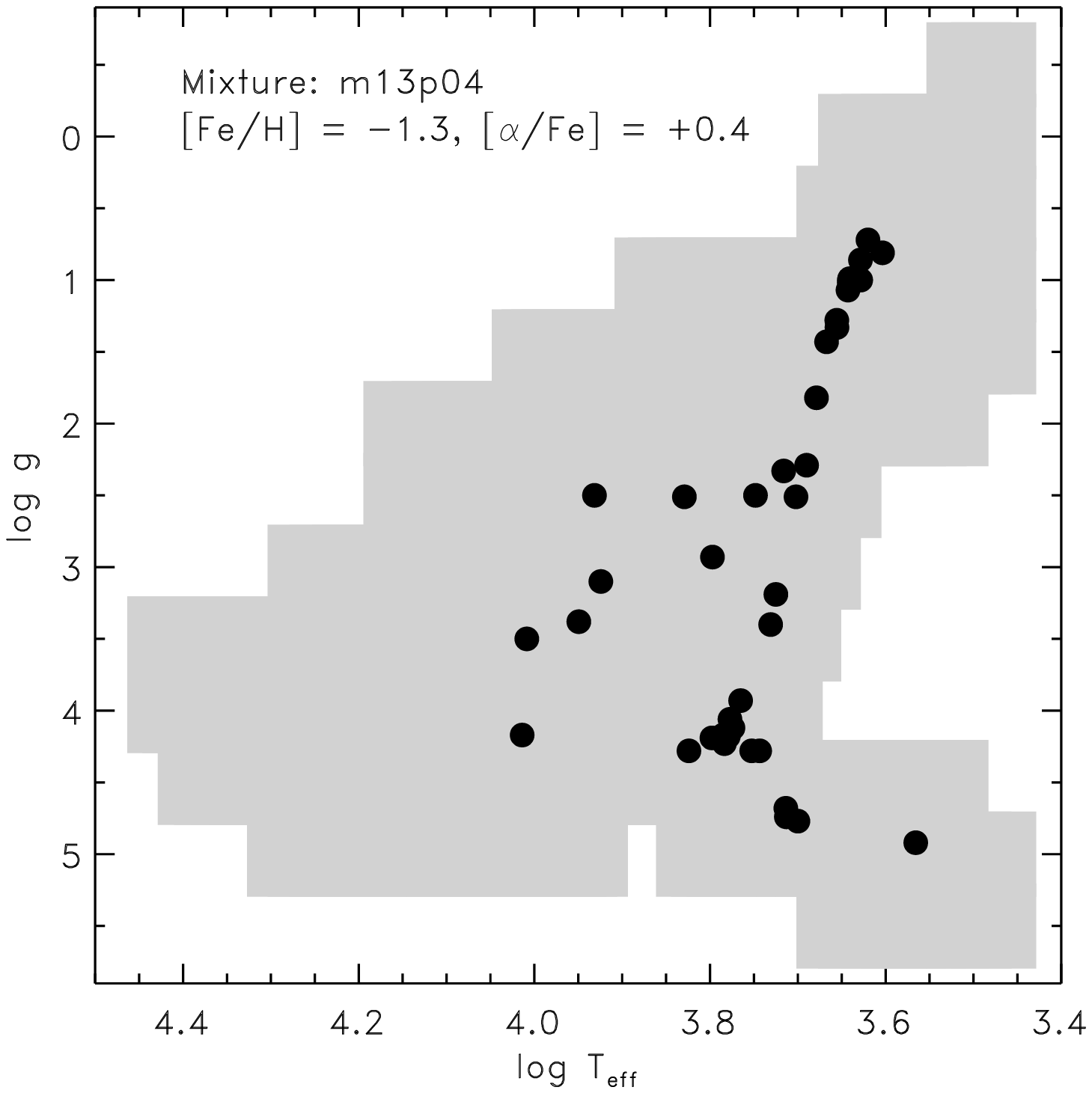}
\includegraphics[width=0.9\columnwidth]{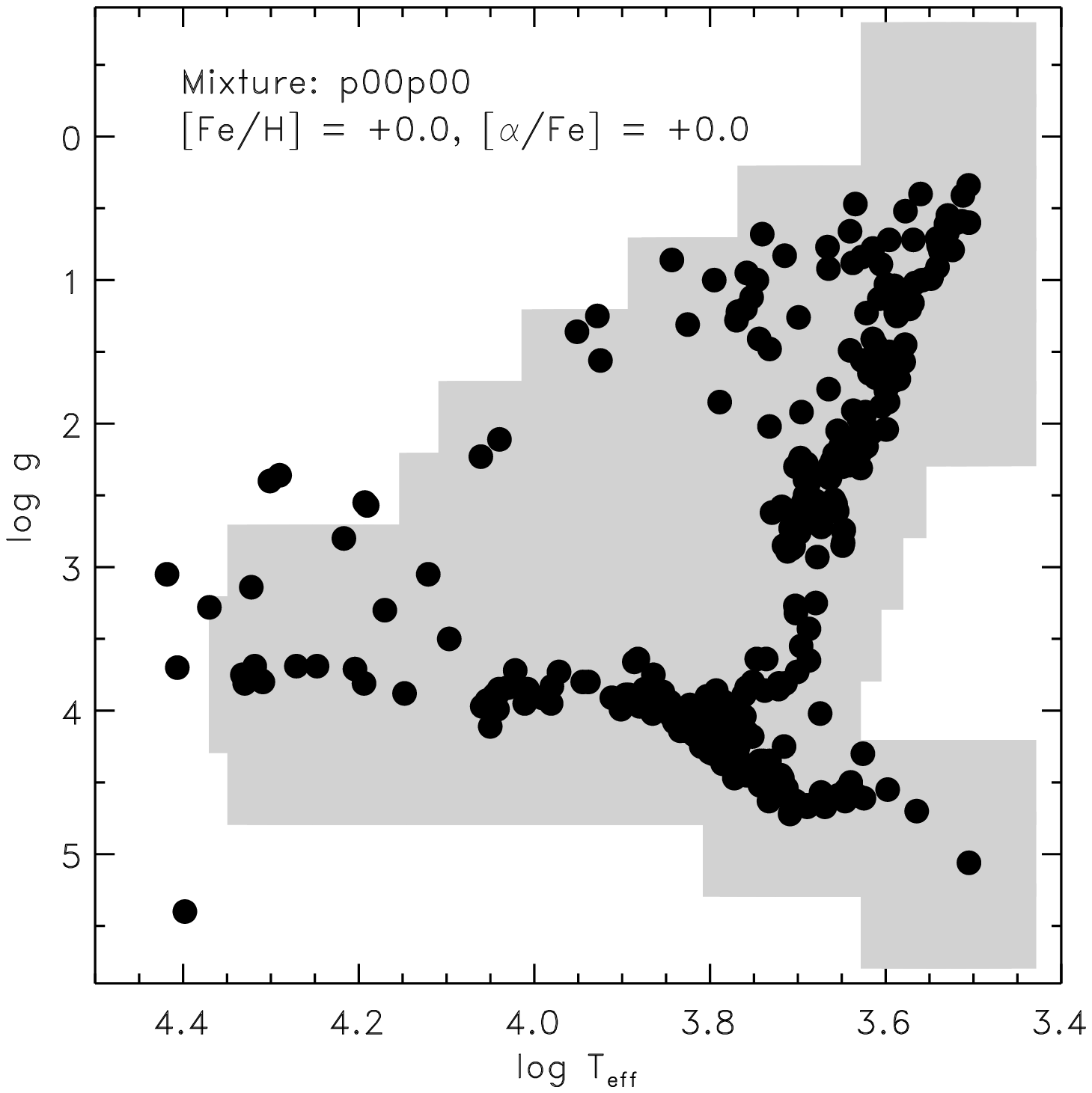}
\caption{The coverage of the model library presented in this work is illustrated by the gray area, in the plane log(\teff) vs. \logg. The coverage of the MILES empirical library is shown as filled circles, adopting the atmospheric parameters derived in \citet{prugniel+11}. Two mixtures are shown, as indicated in the panels.}
\label{fig:coverage_miles}
\end{center}
\end{figure*}

As a second step, the coverage of each selected mixture in \mix{HIGHRES} library were compared to the corresponding MILES coverage in the {\teff} vs. {\logg} space. The comparison for mixtures \mix{m13p04} and \mix{p00p00} are shown in Fig. \ref{fig:coverage_miles}, and the remaining mixtures are shown in the online Appendix A). The grey areas illustrate the coverage of the theoretical library, equally spaced in {\teff} and {\logg} as given in \S\ref{s:lib}. The black filled circles illustrate the stars existing in MILES.
It is important to remember that there is no star in the theoretical library which was not required by stellar evolutionary tracks between 30\,Myr and 14\,Gyr. Therefore, the first thing to notice is that stellar population models computed solely based on MILES will necessarily be extrapolating some regions of the {\teff} vs. {\logg} space. The effect of this uncovered regions on spectral population models will be the subject of a future paper. 

All stars in the selected \mix{HIGHRES} mixtures were fitted in ULySS, but the synthetic spectra which do not have a neighbouring empirical star (where the differences between model and MILES parameters were $\Delta${\teff} $\ge$ 5\%, $\Delta${\logg} $\ge$ 0.3\,dex, $\Delta${\feh} $\ge$ 0.15\,dex) 
were flagged for further identification. The remaining model stars (those which have neighbouring empirical stars) were considered \emph{safe fits}. The \emph{safe fits} correspond to 29\% of the total number of comparisons performed (459 out of 1585).

\begin{figure*}
\begin{center}
\includegraphics[bb=45 60 730 540,width=1.8\columnwidth]{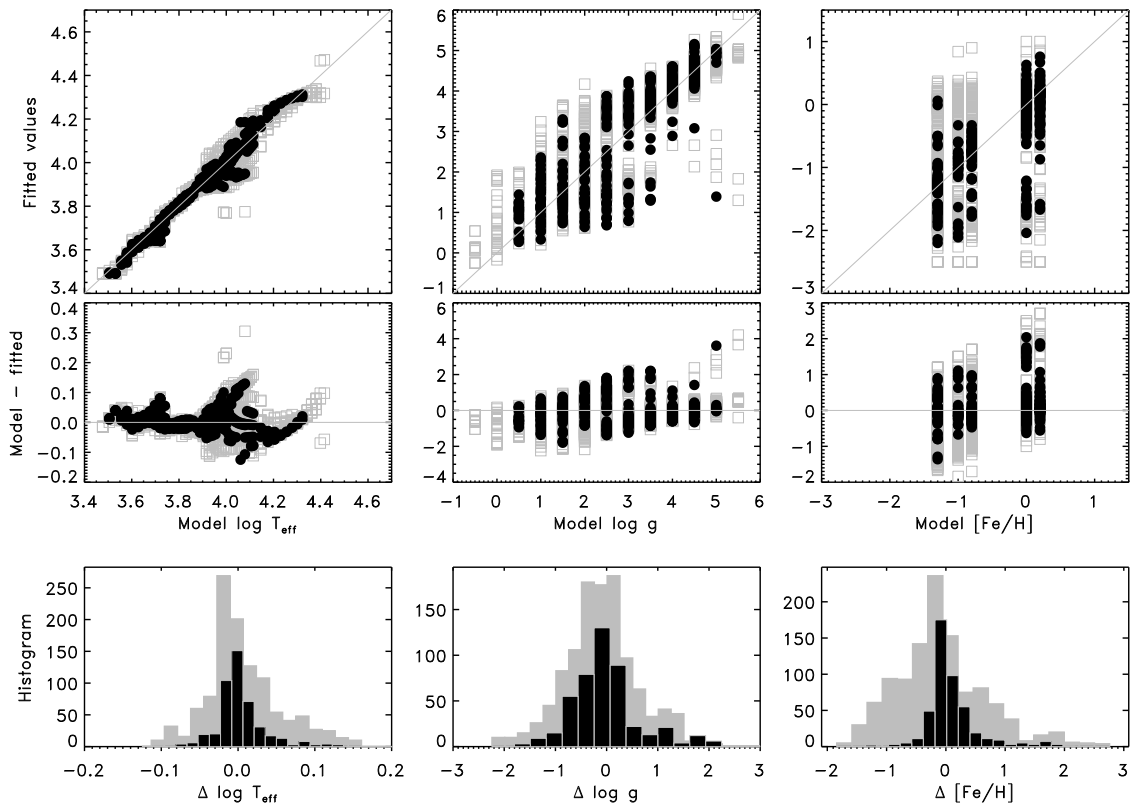}
\caption{Comparisons between atmospheric parameters fitted by \textsc{ULySS} versus the input stellar parameters for all model stars in mixtures \mix{m13p04}, \mix{m08p04}, \mix{p00p00} and \mix{p02p00}. In all panels, the black symbols indicate the model stars whose parameters are well covered in MILES. The gray symbols indicate poorly sampled regions in MILES coverage (where fitted values were either extrapolated or interpolated in regions devoid of stars; see Fig. \ref{fig:coverage_miles}). The left-hand column show results for \teff, the middle column show results for \logg, and the right-hand column show results for [Fe/H]. Input vs. fitted values are shown in the top panels, with residuals shown in the middle row panels. The bottom panels show the histogram distributions of the residuals, computed with bins of 300\,K, 0.25\,dex and 0.15\,dex for {\teff}, {\logg} and [Fe/H] respectively. These results correspond to the fitting  performed in the wavelength range  4200 - 6800\,{\AA}.}
\label{fig_ulycomp}
\end{center}
\end{figure*}

Finally, the \mix{HIGHRES} library was fitted for two different wavelength ranges: \emph{(a)} 4200 - 6800\,{\AA}, the same range adopted in \citet{prugniel+11}, and; \emph{(b)} 4828 -- 5364\,{\AA}, the suggested range in \citet{walcher+09} to derive stellar population parameters from integrated spectra. The motivation to study at least two wavelength ranges comes from current evidence that the choice of wavelength range has an impact on the parameters derived in stellar population studies \citep[e.g.][]{walcher+09,cezario+13}. For the fitting process, the recipe delineated in \citet{prugniel+11} was followed, with few modifications: after convolving each synthetic spectra to a spectral resolution of FWHM $\sim$ 2.5\,{\AA} \citep{fbarroso+11},
the spectral fitting was run starting from different guesses, to avoid trapping in local minima. 
The nodes of the starting guesses are the same as in \citet{prugniel+11}: 	
	{\teff} = [3500, 4000, 5600, 7000, 10000, 18000, 30000],
	{\logg} = [1.8, 3.8] and
	{\feh} = [-1.7, -0.3, 0.5]).

The results are shown in Fig. \ref{fig_ulycomp} for the first wavelength range, and in the Appendix for the second wavelength range. The figures show the model versus fitted parameters and corresponding residuals (first and second rows, respectively). The third row shows the histogram distributions of the residuals. The columns, from left to right, shows results for {\teff}, {\logg} and [Fe/H] respectively. In all panels, black symbols correspond to \emph{safe fits}, i.e., within close coverage of MILES library, as defined above. Grey symbols are the stars in regions not well covered by MILES (they were either extrapolated by the spectral interpolator or interpolated in  regions devoid of stars, such as the central void seen in the right-hand panel in Fig. \ref{fig:coverage_miles}).

At first, one notices that the distribution of residuals between safe and unsafe fits (black and grey symbols) can be notably different. This raises a warning of caution over using spectral interpolator beyond the close coverage of the library where it was derived from.  
Secondly, 
the histograms of residuals are centred close to 0, thus zero-points below the uncertainties reported in Table \ref{tab:error}. The mean, median and mean absolute deviations values are shown in Table \ref{tab_ulystats} for both wavelength ranges fitted (only \emph{safe fits} considered). 

\begin{table}
  \centering
\caption{\label{tab_ulystats} Mean, median and mean absolute deviation $\overline\Delta$ values of the distributions $\Delta$\teff, $\Delta${\logg} and 
$\Delta${\feh} from Fig. \ref{fig_ulycomp}. Ranges 1 and 2 correspond to fitting performed in the wavelength ranges 4200 -- 6800 and 4828 -- 5364\,{\AA} respectively (see text for details).}
\begin{tabular}{lrr}
\hline
 & Range 1 & Range 2 \\
 \hline
{\teff} (K)\\
Mean              &   --34   &   54 \\
Median            &   --49   &   82 \\
$\overline\Delta$ &   412   &   387 \\
\hline
{\logg} (dex)\\
Mean    & --0.02 & --0.01 \\
Median  & --0.11 &  0.01 \\
$\overline\Delta$ & 0.48 & 0.43 \\
\hline
{\feh} (dex)\\
Mean   & 0.10 & 0.05 \\
Median & 0.00 & --0.01 \\
$\overline\Delta$ & 0.28 & 0.25 \\
\hline
\end{tabular}
\end{table}

\begin{figure}
\begin{center}
\includegraphics[bb=30 30 455 460,width=0.8\columnwidth]{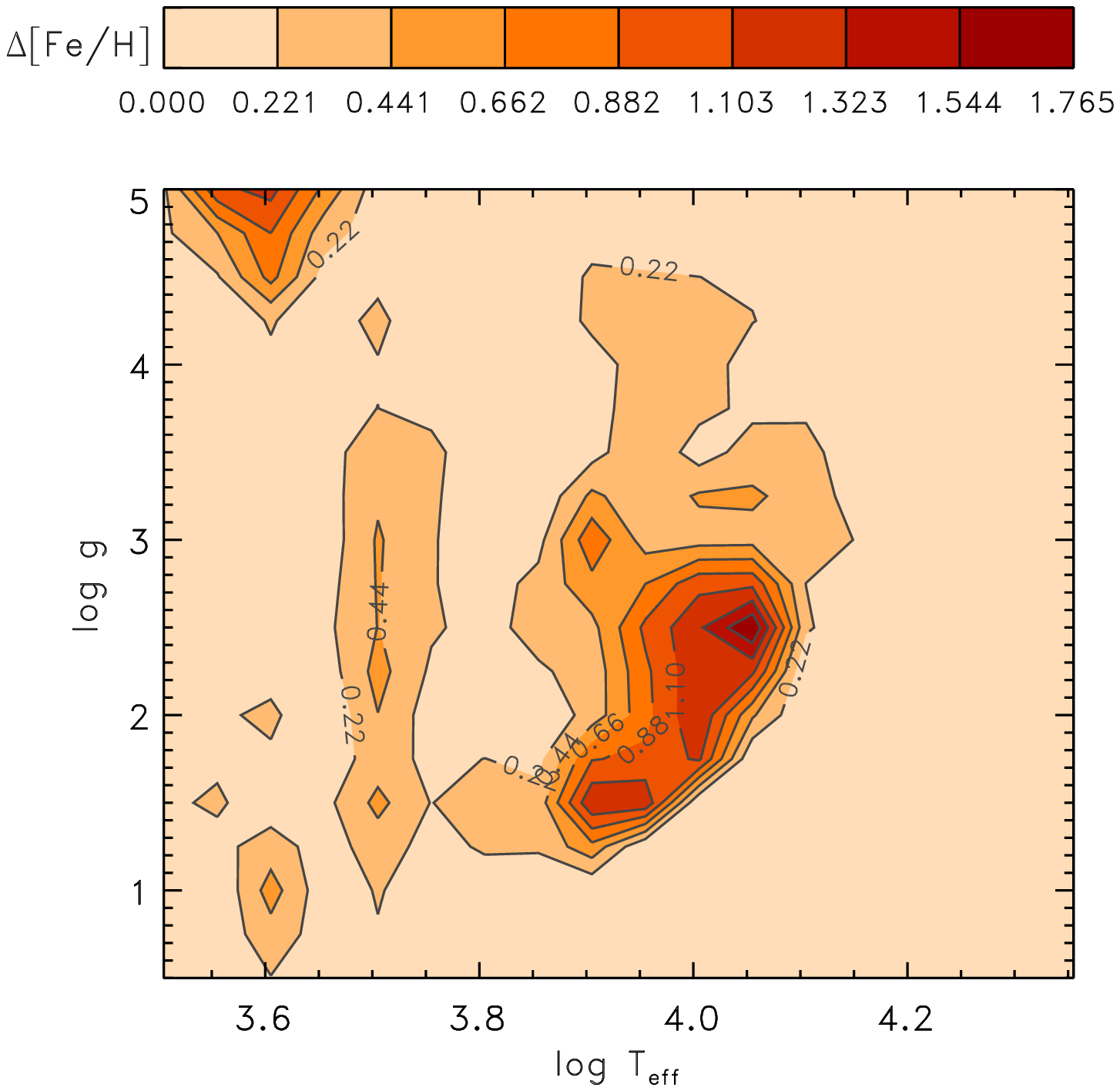}
\includegraphics[bb=30 30 455 460,width=0.8\columnwidth]{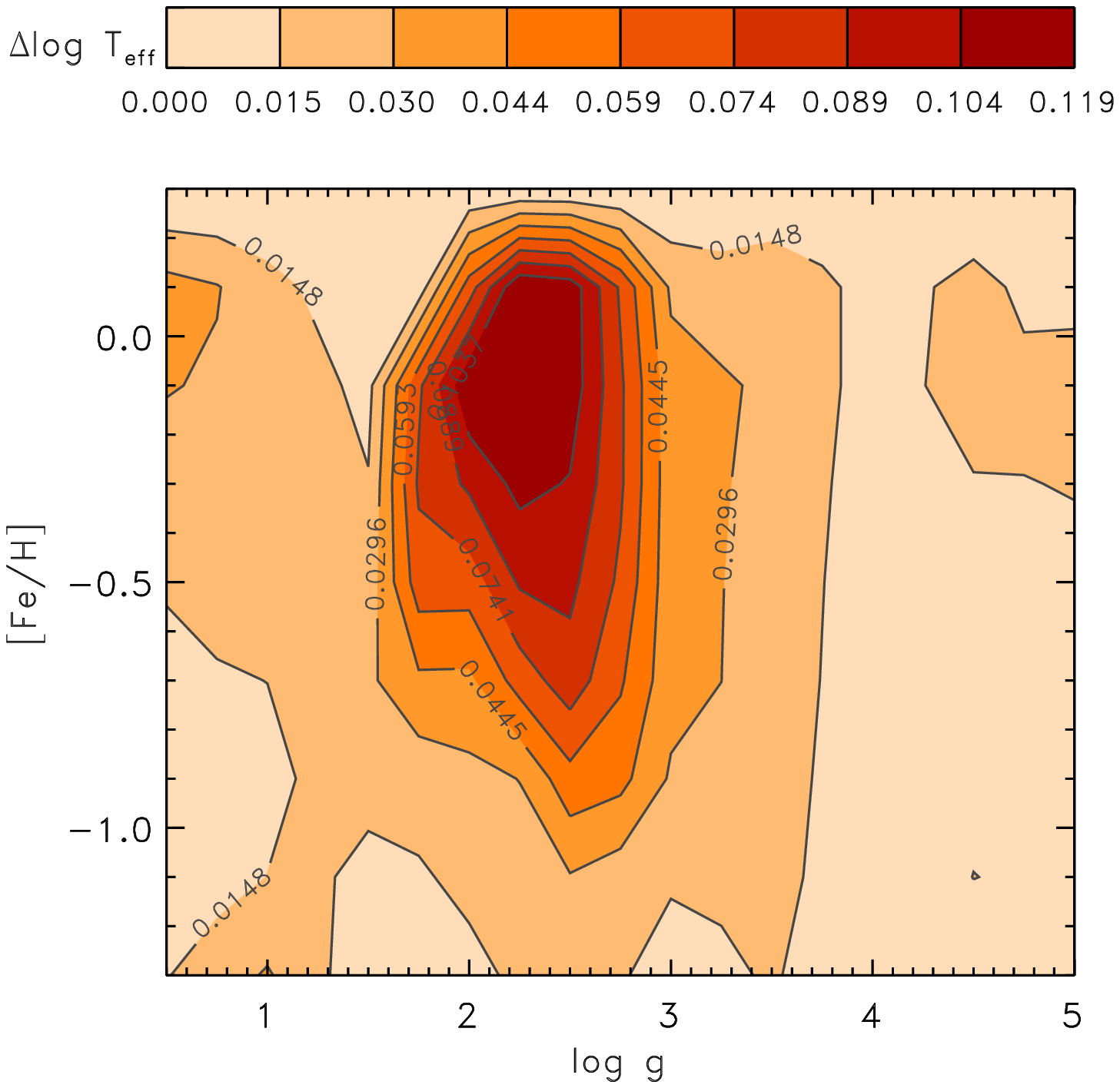}
\includegraphics[bb=30 30 455 460,width=0.8\columnwidth]{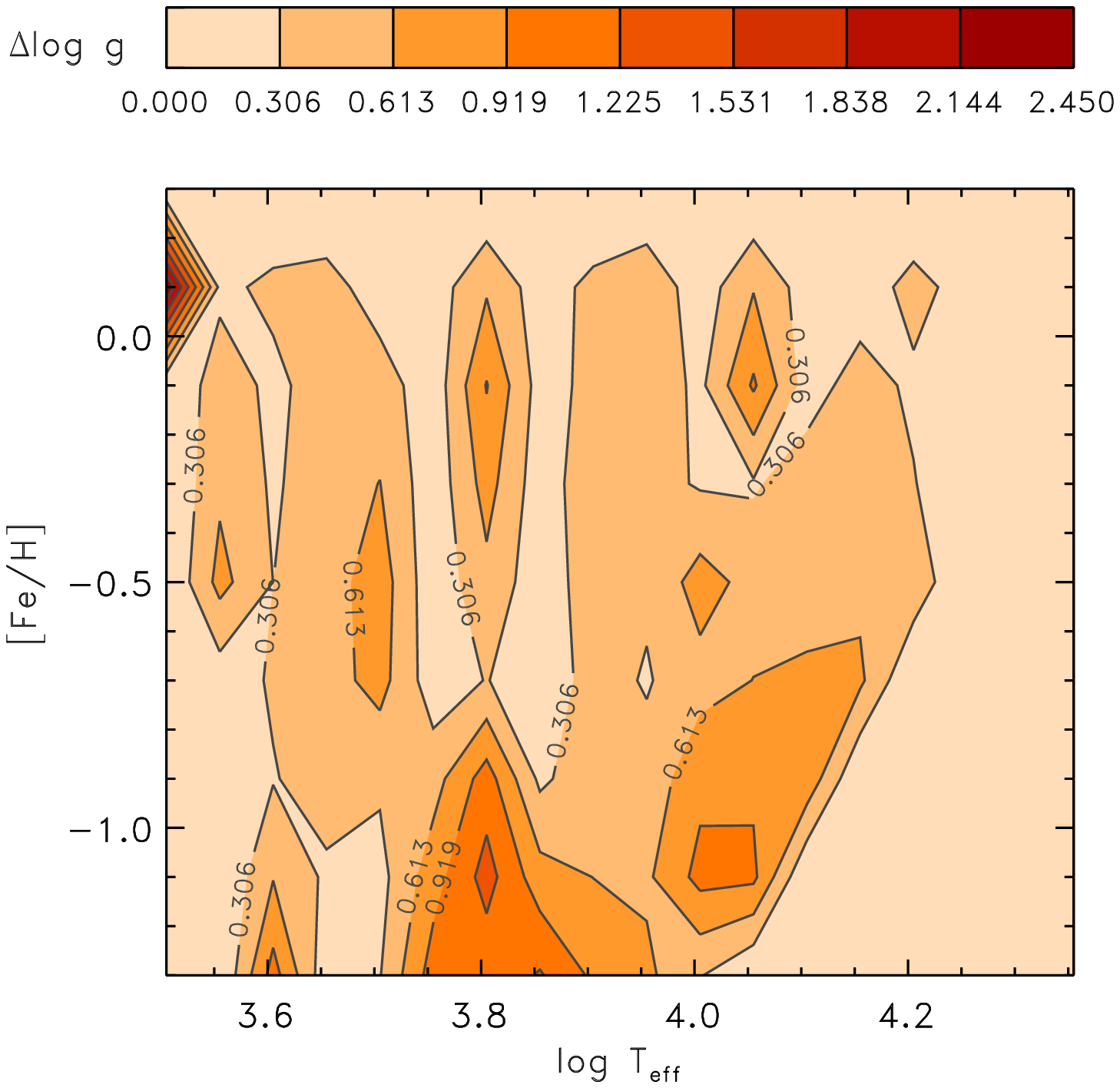}
\caption{The absolute differences between model and fitted parameters are presented as contour plots. Fitted values correspond to parameters derived for the models from a spectral interpolator based on the empirical library MILES. The top panel shows the absolute $\Delta$[Fe/H] in the plane log {\teff} vs. \logg. The middle row panel shows the absolute $\Delta$ log {\teff} in the plane {\logg} vs. [Fe/H]. The bottom panel shows the absolute $\Delta$ {\logg} in the plane log {\teff} vs. [Fe/H]. }
\label{fig:resmapfeh}
\end{center}
\end{figure}

There are few cases with very deviant model vs. fitted parameter. In order to locate these cases, the absolute differences between model and fitted parameters  
and shown in Fig. \ref{fig:resmapfeh} as contour plots. The top panel shows the contour regions of $|\Delta${[Fe/H]}$|$ in the $\log$({\teff}) vs. {\logg} space. It can be seen that the deviant regions are cool giants and a region centred at [{\teff}, {\logg}] $\sim$ [12500\,K, 2.5 dex]. This last region correspond to stars with relatively weak metal signal, and where the coverage density in MILES is low, often with only one observed star per parameters interval (see Fig. \ref{fig:coverage_miles}). 

The middle panel in Fig. \ref{fig:resmapfeh} shows the contour regions of $|\Delta\log$({\teff})$|$ in the {\logg} vs. [Fe/H] space. The regions of larger deviations are relatively evenly spread over [Fe/H],  occupying mainly the region 2.0 $\apprle$ {\logg} $\apprle$ 3.0. By looking at the middle row in Fig. \ref{fig_ulycomp}, it is noticeable a feature with larger deviations around {\teff} $\sim$ 5000\,K. The origin of this pattern is, at the moment, unknown. In advance to compute future model libraries, it could be interesting to investigate in more detail the stars in these regions and identify if the mismatches are related to a characteristic flaw in the models or in the atmospheric parameters adopted for the empirical spectra. Also, larger deviations than average are found for some stars hotter than 12500\,K where, as noted previously, the sampling of stars in MILES is low.
The bottom panel in Fig. \ref{fig:resmapfeh} shows the contour regions of $|\Delta$(\logg)$|$ in the $\log$(\teff) vs. [Fe/H] space. The most deviant region correspond to the coolest regime, which were already shown to be the most deviating model spectra (Table \ref{t:milespar} and Fig. \ref{fig_miles}).  

\vspace{1cm}

In summary, model spectra were compared to empirical spectra in two ways: by the statistical comparison of the fluxes and by comparing their atmospheric parameters scales.  
Regarding the first comparison, 
it was shown that at medium spectral resolution used nowadays at stellar population studies, the model spectra reproduce the observations for a large range of temperatures and wavelength intervals (typically within 3\% in flux above {\teff} = 6000\,K, raising to 10\% at {\teff} = 4000\,K). 
There are few wavelength regions which deviate and should be further investigated. Regarding the second comparison, on average the parameter scales of the model and empirical library agree within the uncertainties. Among the few deviant cases, at least some of them can be related to the sparsity of the empirical library coverage.

\section{Conclusions}
\label{conclusions}

A library of theoretical stellar spectra is presented. This library consists of newly computed ODFs, ATLAS9 model atmospheres, low resolution fluxes from UV to far--IR (\mix{SED}) and high resolution spectra from 2500 to 9000\,{\AA} (\mix{HIGHRES}). The library comprises 12 chemical abundance mixtures, four of those being scaled-solar and the remaining being enhanced in $\alpha$-elements by 0.4\,dex. 

The intended main use of this library is as an ingredient to fully theoretical and differential stellar population models, therefore the coverage in the {\teff} vs. {\logg} parameters space was fine-tuned to the requirements of stellar evolutionary tracks of populations between 30\,Myr and 14\,Gyr, Z between 0.0017 and 0.049 at both scaled-solar and $\alpha$-enhanced mixtures.

Through the comparison between \mix{SED} and \mix{HIGHRES} models, a study on the spectrophotometric effect of lines with predicted energy levels was performed. It is demonstrated that these lines affect the spectrophotometric predictions for stars below {\teff} $\sim$ 7000\,K only. Ad-hoc correction functions were derived on a star-by-star basis to bring the spectrophotometric predictions of the \mix{HIGHRES} models (which do not include lines with predicted energy levels) into better agreement with those of the \mix{SED} models (which includes them). 

Broad band colours predictions from the \mix{SED} models were compared to a recent empirical {\teff}-colour calibration from literature \citep{worthey_lee11}. Averaged colour differences were derived to be: 
$\Delta$(U--B) =   0.168, 
$\Delta$(B--V) =  -0.032,  
$\Delta$(V--I) =  -0.011, 
$\Delta$(V--R) =  -0.031, 
$\Delta$(J--K) =  -0.018 and
$\Delta$(H--K) =  -0.015.

The \mix{HIGHRES} models were compared to the empirical library MILES \citep{MILES1,MILES2} in two ways: by the  comparison of fluxes and by the comparison of their atmospheric parameters scales. 

For the first test, statistically meaningful comparisons were attempted, though relatively few nodes in {\teff}, {\logg} of the theoretical library  have a counterpart in MILES with a large number of stars. For the solar metalicity, where MILES coverage is best, only 11 pairs [{\teff}, {\logg}] were found with at least eight empirical spectra. A larger number of empirical stars per parameter interval would be desirable, in order to clearly identify where atomic and molecular line lists are systematically deviating from observations. Within the sample studied in this work, there is evidence for: missing opacity below 4200\,{\AA} and excessive opacity in regions dominated by CH A--X, C$_2$ and MgH. In the latter case, it would be interesting to further investigate if the effects at CH A--X and C$_2$ bands could be related to non-solar abundances of C and N. It is also worth remembering that the core of very strong lines, such as H lines in stars hotter than {\teff} = 6500\,K cannot be well reproduced with LTE spectral synthesis and photosphere models alone, and should therefore, be masked in automatic spectral fitting techniques involving model spectrum. 

As a second test of the \mix{HIGHRES} models, atmospheric parameters of the model spectra were compared to parameters derived by a spectral interpolator based on MILES stars \citep{prugniel+11}. No significant systematic difference between model and empirical scales are found, with average differences comfortably below the uncertainties. Some deviant regions are found, and some of those are related to regions of the {\teff} vs. {\logg} plane which are poorly populated in the empirical library. 

This result is particularly reassuring for methods which employ model libraries to automatically derive atmospheric parameters of stars in spectroscopic surveys. These comparisons also highlighted the advantage of model libraries in terms of covering the HR diagram, even when compared to the relatively recent and widely used empirical library MILES. 

\section*{Acknowledgments}
The author would like to thank Lucimara P. Martins for the countless discussions (on the present and several other projects) and the companionship,  Jorge Mel\'endez and Philippe Prugniel for several related discussions, and Carlos Rodrigo Blanco for his assistance making the files public through VO. The author is in debt with Fiorella Castelli, Piercarlo Bonifacio and the Kurucz-discuss mailing list (\url{kurucz-discuss@list.fmf.uni-lj.si}) for years of help regarding Robert Kurucz's codes and data. 
The figures in this manuscript made use of several IDL routines from the Coyote (\url{http://www.idlcoyote.com/}) and NASA Astronomy User's (\url{http://idlastro.gsfc.nasa.gov/}) libraries. This work had financial support from FAPESP (projects 08/58406-4, 09/09465-0 and 13/17216-6) and CNPq (project 304291/2012-9).


\begin{thebibliography}{}

\bibitem[\protect\citeauthoryear{{Asplund}, {Grevesse}, {Sauval} \&
  {Scott}}{{Asplund} et~al.}{2009}]{asplund+09}
{Asplund} M.,  {Grevesse} N.,  {Sauval} A.~J.,    {Scott} P.,  2009, \araa, 47,
  481

\bibitem[\protect\citeauthoryear{{B{\'a}nyai}, {Kiss}, {Bedding}, {Bellamy},
  {Benk{\H o}}, {B{\'o}di} \& {Callingham}}{{B{\'a}nyai}
  et~al.}{2013}]{banyai+13}
{B{\'a}nyai} E.,  {Kiss} L.~L.,  {Bedding} T.~R.,  {Bellamy} B.,  {Benk{\H o}}
  J.~M.,  {B{\'o}di} A.,    {Callingham} J.~R.,  2013, \mnras, 436, 1576

\bibitem[\protect\citeauthoryear{Barbuy, Perrin, Katz, Coelho, Cayrel, Spite \&
  Van't Veer-Menneret}{Barbuy et~al.}{2003}]{barbuy+03}
Barbuy B.,  Perrin M.-N.,  Katz D.,  Coelho P.,  Cayrel R.,  Spite M.,    Van't
  Veer-Menneret C.,  2003, A\&A, 404, 661

\bibitem[\protect\citeauthoryear{{Bell}, {Paltoglou} \& {Tripicco}}{{Bell}
  et~al.}{1994}]{bell+94}
{Bell} R.~A.,  {Paltoglou} G.,    {Tripicco} M.~J.,  1994, MNRAS, 268, 771

\bibitem[\protect\citeauthoryear{{Bertone}, {Buzzoni}, {Ch{\'a}vez} \&
  {Rodr{\'{\i}}guez-Merino}}{{Bertone} et~al.}{2008}]{bertone+08}
{Bertone} E.,  {Buzzoni} A.,  {Ch{\'a}vez} M.,    {Rodr{\'{\i}}guez-Merino}
  L.~H.,  2008, \aap, 485, 823

\bibitem[\protect\citeauthoryear{{Bessell}, {Castelli} \& {Plez}}{{Bessell}
  et~al.}{1998}]{bessell+98}
{Bessell} M.~S.,  {Castelli} F.,    {Plez} B.,  1998, \aap, 333, 231

\bibitem[\protect\citeauthoryear{{Brott} \& {Hauschildt}}{{Brott} \&
  {Hauschildt}}{2005}]{PHOENIX05}
{Brott} I.,  {Hauschildt} P.~H.,  2005, in {Turon} C.,  {O'Flaherty} K.~S.,
  {Perryman} M.~A.~C.,  eds, The Three-Dimensional Universe with Gaia Vol.~576
  of ESA Special Publication, {A PHOENIX Model Atmosphere Grid for Gaia}.
p.~565

\bibitem[\protect\citeauthoryear{{Bruzual} \& {Charlot}}{{Bruzual} \&
  {Charlot}}{2003}]{BC03}
{Bruzual} G.,  {Charlot} S.,  2003, MNRAS, 344, 1000

\bibitem[\protect\citeauthoryear{{Buzzoni}, {Bertone}, {Ch{\'a}vez} \&
  {Rodr{\'{\i}}guez-Merino}}{{Buzzoni} et~al.}{2009}]{buzzoni+09proc}
{Buzzoni} A.,  {Bertone} E.,  {Ch{\'a}vez} M.,    {Rodr{\'{\i}}guez-Merino}
  L.~H.,  2009, in {Ch{\'a}vez Dagostino} M.,  {Bertone} E.,  {Rosa Gonzalez}
  D.,   {Rodriguez-Merino} L.~H.,  eds, New Quests in Stellar Astrophysics. II.
  Ultraviolet Properties of Evolved Stellar Populations {Population Synthesis
  at Short Wavelengths and Spectrophotometric Diagnostic Tools for Galaxy
  Evolution}.
pp 263--271

\bibitem[\protect\citeauthoryear{{Caffau}, {Ludwig}, {Steffen}, {Freytag} \&
  {Bonifacio}}{{Caffau} et~al.}{2011}]{caffau+11}
{Caffau} E.,  {Ludwig} H.-G.,  {Steffen} M.,  {Freytag} B.,    {Bonifacio} P.,
  2011, \solphys, 268, 255

\bibitem[\protect\citeauthoryear{{Cappellari} \& {Emsellem}}{{Cappellari} \&
  {Emsellem}}{2004}]{pPXF}
{Cappellari} M.,  {Emsellem} E.,  2004, \pasp, 116, 138

\bibitem[\protect\citeauthoryear{{Castelli}}{{Castelli}}{2005}]{DFSYNTHE}
{Castelli} F.,  2005, Memorie della Societa Astronomica Italiana Supplement, 8,
  34

\bibitem[\protect\citeauthoryear{{Castelli} \& {Hubrig}}{{Castelli} \&
  {Hubrig}}{2004}]{castelli_hubrig04}
{Castelli} F.,  {Hubrig} S.,  2004, \aap, 425, 263

\bibitem[\protect\citeauthoryear{{Castelli} \& {Kurucz}}{{Castelli} \&
  {Kurucz}}{1994}]{castelli_kurucz94}
{Castelli} F.,  {Kurucz} R.~L.,  1994, \aap, 281, 817

\bibitem[\protect\citeauthoryear{Castelli \& Kurucz}{Castelli \&
  Kurucz}{2003}]{ATLASODFNEW}
Castelli F.,  Kurucz R.~L.,  2003, in Modelling of Stellar Atmospheres No.~210
  in IAU Symp, New grids of atlas9 model atmospheres.
Astronomical Society of the Pacific, p.~A20

\bibitem[\protect\citeauthoryear{{Castelli} \& {Kurucz}}{{Castelli} \&
  {Kurucz}}{2004}]{castelli_kurucz04}
{Castelli} F.,  {Kurucz} R.~L.,  2004, \aap, 419, 725

\bibitem[\protect\citeauthoryear{{Cayrel}}{{Cayrel}}{2002}]{cayrel02}
{Cayrel} R.,  2002, in {Lejeune} T.,  {Fernandes} J.,  eds, Observed HR
  Diagrams and Stellar Evolution Vol.~274 of Astronomical Society of the
  Pacific Conference Series, {Determination of Fundamental Parameters}.
p.~133

\bibitem[\protect\citeauthoryear{{Cenarro}, {Peletier},
  {S{\'a}nchez-Bl{\'a}zquez}, {Selam}, {Toloba}, {Cardiel},
  {Falc{\'o}n-Barroso}, {Gorgas}, {Jim{\'e}nez-Vicente} \&
  {Vazdekis}}{{Cenarro} et~al.}{2007}]{MILES2}
{Cenarro} A.~J.,  {Peletier} R.~F.,  {S{\'a}nchez-Bl{\'a}zquez} P.,  {Selam}
  S.~O.,  {Toloba} E.,  {Cardiel} N.,  {Falc{\'o}n-Barroso} J.,  {Gorgas} J.,
  {Jim{\'e}nez-Vicente} J.,    {Vazdekis} A.,  2007, \mnras, 374, 664

\bibitem[\protect\citeauthoryear{{Cervantes}, {Coelho}, {Barbuy} \&
  {Vazdekis}}{{Cervantes} et~al.}{2007}]{cervantes+07proc}
{Cervantes} J.~L.,  {Coelho} P.,  {Barbuy} B.,    {Vazdekis} A.,  2007, in
  {Vazdekis} A.,  {Peletier} R.~F.,  eds, Stellar Populations as Building
  Blocks of Galaxies Vol.~241 of IAU Symp, {A new approach to derive
  [{$\alpha$}/Fe] for integrated stellar populations}.
Cambridge University Press, pp 167--168

\bibitem[\protect\citeauthoryear{{Cezario}, {Coelho}, {Alves-Brito}, {Forbes}
  \& {Brodie}}{{Cezario} et~al.}{2013}]{cezario+13}
{Cezario} E.,  {Coelho} P.~R.~T.,  {Alves-Brito} A.,  {Forbes} D.~A.,
  {Brodie} J.~P.,  2013, \aap, 549, A60

\bibitem[\protect\citeauthoryear{{Cid Fernandes}, {Mateus}, {Sodr{\'e}},
  {Stasi{\'n}ska} \& {Gomes}}{{Cid Fernandes} et~al.}{2005}]{cid+05}
{Cid Fernandes} R.,  {Mateus} A.,  {Sodr{\'e}} L.,  {Stasi{\'n}ska} G.,
  {Gomes} J.~M.,  2005, MNRAS, 358, 363

\bibitem[\protect\citeauthoryear{{Code}, {Bless}, {Davis} \& {Brown}}{{Code}
  et~al.}{1976}]{code+76}
{Code} A.~D.,  {Bless} R.~C.,  {Davis} J.,    {Brown} R.~H.,  1976, \apj, 203,
  417

\bibitem[\protect\citeauthoryear{{Coelho}}{{Coelho}}{2009}]{coelho09a_proc}
{Coelho} P.,  2009, in {Giobbi} G.,  {Tornambe} A.,  {Raimondo} G.,  {Limongi}
  M.,  {Antonelli} L.~A.,  {Menci} N.,   {Brocato} E.,  eds, American Institute
  of Physics Conference Series Vol.~1111 of American Institute of Physics
  Conference Series, {Spectral libraries and their uncertainties}.
pp 67--74

\bibitem[\protect\citeauthoryear{Coelho, Barbuy, Melendez, Schiavon \&
  Castilho}{Coelho et~al.}{2005}]{coelho+05}
Coelho P.,  Barbuy B.,  Melendez J.,  Schiavon R.,    Castilho B.,  2005, A\&A,
  443, 735

\bibitem[\protect\citeauthoryear{{Coelho}, {Bruzual}, {Charlot}, {Weiss},
  {Barbuy} \& {Ferguson}}{{Coelho} et~al.}{2007}]{coelho+07}
{Coelho} P.,  {Bruzual} G.,  {Charlot} S.,  {Weiss} A.,  {Barbuy} B.,
  {Ferguson} J.~W.,  2007, \mnras, 382, 498

\bibitem[\protect\citeauthoryear{{Conroy} \& {van Dokkum}}{{Conroy} \& {van
  Dokkum}}{2012}]{conroy_dokkum12}
{Conroy} C.,  {van Dokkum} P.,  2012, \apj, 747, 69

\bibitem[\protect\citeauthoryear{{de Laverny}, {Recio-Blanco}, {Worley} \&
  {Plez}}{{de Laverny} et~al.}{2012}]{delaverny+12}
{de Laverny} P.,  {Recio-Blanco} A.,  {Worley} C.~C.,    {Plez} B.,  2012,
  \aap, 544, A126

\bibitem[\protect\citeauthoryear{{Delgado}, {Cervi{\~n}o}, {Martins},
  {Leitherer} \& {Hauschildt}}{{Delgado} et~al.}{2005}]{delgado+05}
{Delgado} R.~M.~G.,  {Cervi{\~n}o} M.,  {Martins} L.~P.,  {Leitherer} C.,
  {Hauschildt} P.~H.,  2005, MNRAS, 357, 945

\bibitem[\protect\citeauthoryear{{di Benedetto}}{{di
  Benedetto}}{1993}]{dibenedetto93}
{di Benedetto} G.~P.,  1993, \aap, 270, 315

\bibitem[\protect\citeauthoryear{{Dyck}, {Benson}, {van Belle} \&
  {Ridgway}}{{Dyck} et~al.}{1996}]{dyck+96}
{Dyck} H.~M.,  {Benson} J.~A.,  {van Belle} G.~T.,    {Ridgway} S.~T.,  1996,
  \aj, 111, 1705

\bibitem[\protect\citeauthoryear{{Falc{\'o}n-Barroso},
  {S{\'a}nchez-Bl{\'a}zquez}, {Vazdekis}, {Ricciardelli}, {Cardiel}, {Cenarro},
  {Gorgas} \& {Peletier}}{{Falc{\'o}n-Barroso} et~al.}{2011}]{fbarroso+11}
{Falc{\'o}n-Barroso} J.,  {S{\'a}nchez-Bl{\'a}zquez} P.,  {Vazdekis} A.,
  {Ricciardelli} E.,  {Cardiel} N.,  {Cenarro} A.~J.,  {Gorgas} J.,
  {Peletier} R.~F.,  2011, \aap, 532, A95

\bibitem[\protect\citeauthoryear{Fr{\'e}maux, Kupka, Boisson, Joly \&
  Tsymbal}{Fr{\'e}maux et~al.}{2006}]{fremaux+06}
Fr{\'e}maux J.,  Kupka F.,  Boisson C.,  Joly M.,    Tsymbal V.,  2006, A\&A,
  449, 109

\bibitem[\protect\citeauthoryear{{Grevesse} \& {Sauval}}{{Grevesse} \&
  {Sauval}}{1998}]{gs98}
{Grevesse} N.,  {Sauval} A.~J.,  1998, \ssr, 85, 161

\bibitem[\protect\citeauthoryear{{Gustafsson}, {Edvardsson}, {Eriksson},
  {Jorgensen}, {Nordlund} \& {Plez}}{{Gustafsson} et~al.}{2008}]{MARCS08}
{Gustafsson} B.,  {Edvardsson} B.,  {Eriksson} K.,  {Jorgensen} U.~G.,
  {Nordlund} {\AA}.,    {Plez} B.,  2008, \aap, 486, 951

\bibitem[\protect\citeauthoryear{{Guti{\'e}rrez}, {Rodrigo} \&
  {Solano}}{{Guti{\'e}rrez} et~al.}{2006}]{gutierrez+06proc}
{Guti{\'e}rrez} R.,  {Rodrigo} C.,    {Solano} E.,  2006, in {Gabriel} C.,
  {Arviset} C.,  {Ponz} D.,   {Enrique} S.,  eds, Astronomical Data Analysis
  Software and Systems XV Vol.~351 of Astronomical Society of the Pacific
  Conference Series, {The Spanish Virtual Observatory (SVO)}.
p.~19

\bibitem[\protect\citeauthoryear{{Heiter} \& {Eriksson}}{{Heiter} \&
  {Eriksson}}{2006}]{heiter_eriksson06}
{Heiter} U.,  {Eriksson} K.,  2006, \aap, 452, 1039

\bibitem[\protect\citeauthoryear{{Jerzykiewicz} \&
  {Molenda-Zakowicz}}{{Jerzykiewicz} \&
  {Molenda-Zakowicz}}{2000}]{jerzykiewicz_zakowicz00}
{Jerzykiewicz} M.,  {Molenda-Zakowicz} J.,  2000, Acta Astronomica, 50, 369

\bibitem[\protect\citeauthoryear{{Johnson} \& {Krupp}}{{Johnson} \&
  {Krupp}}{1976}]{johnson_krupp76}
{Johnson} H.~R.,  {Krupp} B.~M.,  1976, \apj, 206, 201

\bibitem[\protect\citeauthoryear{{Jones} \& {Tsuji}}{{Jones} \&
  {Tsuji}}{1997}]{jones_tsuji97}
{Jones} H.~R.~A.,  {Tsuji} T.,  1997, \apjl, 480, L39

\bibitem[\protect\citeauthoryear{{Katz}, {Soubiran}, {Cayrel}, {Adda} \&
  {Cautain}}{{Katz} et~al.}{1998}]{katz+98}
{Katz} D.,  {Soubiran} C.,  {Cayrel} R.,  {Adda} M.,    {Cautain} R.,  1998,
  \aap, 338, 151

\bibitem[\protect\citeauthoryear{{Kervella}, {Th{\'e}venin}, {Di Folco} \&
  {S{\'e}gransan}}{{Kervella} et~al.}{2004}]{kervella+04}
{Kervella} P.,  {Th{\'e}venin} F.,  {Di Folco} E.,    {S{\'e}gransan} D.,
  2004, \aap, 426, 297

\bibitem[\protect\citeauthoryear{{Kirby}}{{Kirby}}{2011}]{kirby11}
{Kirby} E.~N.,  2011, \pasp, 123, 531

\bibitem[\protect\citeauthoryear{{Koleva}, {Prugniel}, {Bouchard} \&
  {Wu}}{{Koleva} et~al.}{2009}]{ULYSS}
{Koleva} M.,  {Prugniel} P.,  {Bouchard} A.,    {Wu} Y.,  2009, \aap, 501, 1269

\bibitem[\protect\citeauthoryear{{Kurucz}}{{Kurucz}}{1993}]{kurucz93CD15}
{Kurucz} R.,  1993, Diatomic Molecular Data for Opacity Calculations.~Kurucz
  CD-ROM No.~15.~Cambridge, Mass.: Smithsonian Astrophysical Observatory,
  1993., 15

\bibitem[\protect\citeauthoryear{{Kurucz}}{{Kurucz}}{1970}]{ATLAS1970}
{Kurucz} R.~L.,  1970, SAO Special Report, 309

\bibitem[\protect\citeauthoryear{{Kurucz}}{{Kurucz}}{1992}]{kurucz92}
{Kurucz} R.~L.,  1992, Revista Mexicana de Astronomia y Astrofisica, vol.~23,
  23, 45

\bibitem[\protect\citeauthoryear{{Kurucz}}{{Kurucz}}{2005a}]{kurucz_codes05}
{Kurucz} R.~L.,  2005a, Memorie della Societa Astronomica Italiana Supplement,
  8, 14

\bibitem[\protect\citeauthoryear{{Kurucz}}{{Kurucz}}{2005b}]{kurucz05}
{Kurucz} R.~L.,  2005b, Memorie della Societa Astronomica Italiana Supplementi,
  8, 76

\bibitem[\protect\citeauthoryear{{Kurucz}}{{Kurucz}}{2006}]{kurucz06_lines}
{Kurucz} R.~L.,  2006, in {Stee} P.,  ed., Radiative Transfer and Applications
  to Very Large Telescopes Vol.~18 of EAS Publications Series, {Including all
  the Lines}.
pp 129--155

\bibitem[\protect\citeauthoryear{{Kurucz} \& {Avrett}}{{Kurucz} \&
  {Avrett}}{1981}]{kurucz_avrett81}
{Kurucz} R.~L.,  {Avrett} E.~H.,  1981, SAO Special Report, 391

\bibitem[\protect\citeauthoryear{{Kurucz}, {Furenlid}, {Brault} \&
  {Testerman}}{{Kurucz} et~al.}{1984}]{solarflux:kurucz84}
{Kurucz} R.~L.,  {Furenlid} I.,  {Brault} J.,    {Testerman} L.,  1984, {Solar
  flux atlas from 296 to 1300 nm}.
National Solar Observatory Atlas, Sunspot, New Mexico: National Solar
  Observatory, 1984

\bibitem[\protect\citeauthoryear{{Ku{\v c}inskas}, {Hauschildt}, {Ludwig},
  {Brott}, {Vansevi{\v c}ius}, {Lindegren}, {Tanab{\'e}} \& {Allard}}{{Ku{\v
  c}inskas} et~al.}{2005}]{kucinskas+05}
{Ku{\v c}inskas} A.,  {Hauschildt} P.~H.,  {Ludwig} H.-G.,  {Brott} I.,
  {Vansevi{\v c}ius} V.,  {Lindegren} L.,  {Tanab{\'e}} T.,    {Allard} F.,
  2005, \aap, 442, 281

\bibitem[\protect\citeauthoryear{{LeBlanc}}{{LeBlanc}}{2010}]{leblanc2010}
{LeBlanc} F.,  2010, An introduction to stellar astrophysics.
John Wiley and Sons, Ltd.

\bibitem[\protect\citeauthoryear{{Lebzelter}, {Heiter}, {Abia}, {Eriksson},
  {Ireland}, {Neilson}, {Nowotny}, {Maldonado}, {Merle}, {Peterson}, {Plez},
  {Short}, {Wahlgren}, {Worley}, {Aringer} \& et al.}{{Lebzelter}
  et~al.}{2012}]{lebzelter+12}
{Lebzelter} T.,  {Heiter} U.,  {Abia} C.,  {Eriksson} K.,  {Ireland} M.,
  {Neilson} H.,  {Nowotny} W.,  {Maldonado} J.,  {Merle} T.,  {Peterson} R.,
  {Plez} B.,  {Short} C.~I.,  {Wahlgren} G.~M.,  {Worley} C.,  {Aringer} B.,
  et al. 2012, \aap, 547, A108

\bibitem[\protect\citeauthoryear{{Lee}, {Worthey}, {Dotter}, {Chaboyer},
  {Jevremovi{\'c}}, {Baron}, {Briley}, {Ferguson}, {Coelho} \& {Trager}}{{Lee}
  et~al.}{2009}]{lee+09}
{Lee} H.-c.,  {Worthey} G.,  {Dotter} A.,  {Chaboyer} B.,  {Jevremovi{\'c}} D.,
   {Baron} E.,  {Briley} M.~M.,  {Ferguson} J.~W.,  {Coelho} P.,    {Trager}
  S.~C.,  2009, \apj, 694, 902

\bibitem[\protect\citeauthoryear{{Leitherer}, {Ortiz Ot{\'a}lvaro}, {Bresolin},
  {Kudritzki}, {Lo Faro}, {Pauldrach}, {Pettini} \& {Rix}}{{Leitherer}
  et~al.}{2010}]{leitherer+10}
{Leitherer} C.,  {Ortiz Ot{\'a}lvaro} P.~A.,  {Bresolin} F.,  {Kudritzki}
  R.-P.,  {Lo Faro} B.,  {Pauldrach} A.~W.~A.,  {Pettini} M.,    {Rix} S.~A.,
  2010, \apjs, 189, 309

\bibitem[\protect\citeauthoryear{{Leitherer}, {Schaerer}, {Goldader},
  {Delgado}, {Robert}, {Kune}, {de Mello}, {Devost} \& {Heckman}}{{Leitherer}
  et~al.}{1999}]{leitherer+99}
{Leitherer} C.,  {Schaerer} D.,  {Goldader} J.~D.,  {Delgado} R.~M.~G.,
  {Robert} C.,  {Kune} D.~F.,  {de Mello} D.~F.,  {Devost} D.,    {Heckman}
  T.~M.,  1999, \apjs, 123, 3

\bibitem[\protect\citeauthoryear{{Lejeune}, {Cuisinier} \& {Buser}}{{Lejeune}
  et~al.}{1997}]{basel1}
{Lejeune} T.,  {Cuisinier} F.,    {Buser} R.,  1997, A\&AS, 125, 229

\bibitem[\protect\citeauthoryear{{Lejeune}, {Cuisinier} \& {Buser}}{{Lejeune}
  et~al.}{1998}]{basel2}
{Lejeune} T.,  {Cuisinier} F.,    {Buser} R.,  1998, A\&AS, 130, 65

\bibitem[\protect\citeauthoryear{{Maraston} \& {Str{\"o}mb{\"a}ck}}{{Maraston}
  \& {Str{\"o}mb{\"a}ck}}{2011}]{maraston_stromback11}
{Maraston} C.,  {Str{\"o}mb{\"a}ck} G.,  2011, \mnras, 418, 2785

\bibitem[\protect\citeauthoryear{{Markwardt}}{{Markwardt}}{2009}]{MPFIT}
{Markwardt} C.~B.,  2009, in {Bohlender} D.~A.,  {Durand} D.,   {Dowler} P.,
  eds, Astronomical Data Analysis Software and Systems XVIII Vol.~411 of
  Astronomical Society of the Pacific Conference Series, {Non-linear
  Least-squares Fitting in IDL with MPFIT}.
p.~251

\bibitem[\protect\citeauthoryear{{Mart{\'{\i}}n-Hern{\'a}ndez},
  {M{\'a}rmol-Queralt{\'o}}, {Gorgas}, {Cardiel}, {S{\'a}nchez-Bl{\'a}zquez},
  {Cenarro}, {Peletier}, {Vazdekis} \&
  {Falc{\'o}n-Barroso}}{{Mart{\'{\i}}n-Hern{\'a}ndez}
  et~al.}{2010}]{martin-hernandez+10proc}
{Mart{\'{\i}}n-Hern{\'a}ndez} J.~M.,  {M{\'a}rmol-Queralt{\'o}} E.,  {Gorgas}
  J.,  {Cardiel} N.,  {S{\'a}nchez-Bl{\'a}zquez} P.,  {Cenarro} A.~J.,
  {Peletier} R.~F.,  {Vazdekis} A.,    {Falc{\'o}n-Barroso} J.,  2010, in
  {Diego} J.~M.,  {Goicoechea} L.~J.,  {Gonz{\'a}lez-Serrano} J.~I.,   {Gorgas}
  J.,  eds, Highlights of Spanish Astrophysics V {New Empirical Fitting
  Functions for the Lick/IDS Indices Using MILES}.
p.~309

\bibitem[\protect\citeauthoryear{{Martins} \& {Coelho}}{{Martins} \&
  {Coelho}}{2007}]{MC07}
{Martins} L.~P.,  {Coelho} P.,  2007, \mnras, 381, 1329

\bibitem[\protect\citeauthoryear{{Martins}, {Delgado}, {Leitherer},
  {Cervi{\~n}o} \& {Hauschildt}}{{Martins} et~al.}{2005}]{martins+05}
{Martins} L.~P.,  {Delgado} R.~M.~G.,  {Leitherer} C.,  {Cervi{\~n}o} M.,
  {Hauschildt} P.,  2005, MNRAS, 358, 49

\bibitem[\protect\citeauthoryear{{McWilliam}}{{McWilliam}}{1997}]{mcwilliam97}
{McWilliam} A.,  1997, \araa, 35, 503

\bibitem[\protect\citeauthoryear{{M{\'e}sz{\'a}ros}, {Allende Prieto},
  {Edvardsson}, {Castelli}, {Garc{\'{\i}}a P{\'e}rez}, {Gustafsson},
  {Majewski}, {Plez}, {Schiavon}, {Shetrone} \& {de
  Vicente}}{{M{\'e}sz{\'a}ros} et~al.}{2012}]{meszaros+12}
{M{\'e}sz{\'a}ros} S.,  {Allende Prieto} C.,  {Edvardsson} B.,  {Castelli} F.,
  {Garc{\'{\i}}a P{\'e}rez} A.~E.,  {Gustafsson} B.,  {Majewski} S.~R.,  {Plez}
  B.,  {Schiavon} R.,  {Shetrone} M.,    {de Vicente} A.,  2012, \aj, 144, 120

\bibitem[\protect\citeauthoryear{{Milone}, {Sansom} \&
  {S{\'a}nchez-Bl{\'a}zquez}}{{Milone} et~al.}{2011}]{milone+11}
{Milone} A.~D.~C.,  {Sansom} A.~E.,    {S{\'a}nchez-Bl{\'a}zquez} P.,  2011,
  \mnras, 414, 1227

\bibitem[\protect\citeauthoryear{Mor{\'e}}{Mor{\'e}}{1978}]{MINPACK}
Mor{\'e} J.~J.,  1978, in Watson G.~A.,  ed., Numerical Analysis: Lecture Notes
  in Mathematics Vol.~630, The levenberg-marquardt algorithm: Implementation
  and theory.
Springer-Verlag: Berlin, pp 105--116

\bibitem[\protect\citeauthoryear{Munari, Sordo, Castelli \& Zwitter}{Munari
  et~al.}{2005}]{munari+05}
Munari U.,  Sordo R.,  Castelli F.,    Zwitter T.,  2005, A\&A, 442, 1127

\bibitem[\protect\citeauthoryear{{Murphy} \& {Meiksin}}{{Murphy} \&
  {Meiksin}}{2004}]{murphy_meiksin04}
{Murphy} T.,  {Meiksin} A.,  2004, \mnras, 351, 1430

\bibitem[\protect\citeauthoryear{Nyquist}{Nyquist}{2002}]{nyquist}
Nyquist H.,  2002, Proceedings of the IEEE, 90, 280

\bibitem[\protect\citeauthoryear{{Palacios}, {Gebran}, {Josselin}, {Martins},
  {Plez}, {Belmas} \& {L{\`e}bre}}{{Palacios} et~al.}{2010}]{palacios+10}
{Palacios} A.,  {Gebran} M.,  {Josselin} E.,  {Martins} F.,  {Plez} B.,
  {Belmas} M.,    {L{\`e}bre} A.,  2010, \aap, 516, A13

\bibitem[\protect\citeauthoryear{{Partridge} \& {Schwenke}}{{Partridge} \&
  {Schwenke}}{1997}]{partridge_schwenke97}
{Partridge} H.,  {Schwenke} D.~W.,  1997, Journal of Chemical Physics, 106,
  4618

\bibitem[\protect\citeauthoryear{{Pence}, {Chiappetti}, {Page}, {Shaw} \&
  {Stobie}}{{Pence} et~al.}{2010}]{pence+10}
{Pence} W.~D.,  {Chiappetti} L.,  {Page} C.~G.,  {Shaw} R.~A.,    {Stobie} E.,
  2010, \aap, 524, A42

\bibitem[\protect\citeauthoryear{{Percival} \& {Salaris}}{{Percival} \&
  {Salaris}}{2009}]{percival_salaris09}
{Percival} S.~M.,  {Salaris} M.,  2009, \apj, 703, 1123

\bibitem[\protect\citeauthoryear{{Percival}, {Salaris}, {Cassisi} \&
  {Pietrinferni}}{{Percival} et~al.}{2009}]{percival+09}
{Percival} S.~M.,  {Salaris} M.,  {Cassisi} S.,    {Pietrinferni} A.,  2009,
  \apj, 690, 427

\bibitem[\protect\citeauthoryear{{Peterson}, {Dorman} \& {Rood}}{{Peterson}
  et~al.}{2001}]{peterson+01}
{Peterson} R.~C.,  {Dorman} B.,    {Rood} R.~T.,  2001, ApJ, 559, 372

\bibitem[\protect\citeauthoryear{{Pietrinferni}, {Cassisi}, {Salaris} \&
  {Castelli}}{{Pietrinferni} et~al.}{2004}]{pietrinferni+04}
{Pietrinferni} A.,  {Cassisi} S.,  {Salaris} M.,    {Castelli} F.,  2004, \apj,
  612, 168

\bibitem[\protect\citeauthoryear{{Plez}}{{Plez}}{2008}]{plez08proc}
{Plez} B.,  2008, Physica Scripta Volume T, 133, 014003

\bibitem[\protect\citeauthoryear{{Plez}}{{Plez}}{2011}]{plez11proc}
{Plez} B.,  2011, Journal of Physics Conference Series, 328, 012005

\bibitem[\protect\citeauthoryear{{Proctor} \& {Sansom}}{{Proctor} \&
  {Sansom}}{2002}]{proctor_sansom02}
{Proctor} R.~N.,  {Sansom} A.~E.,  2002, MNRAS, 333, 517

\bibitem[\protect\citeauthoryear{{Prugniel}, {Koleva}, {Ocvirk}, {Le Borgne} \&
  {Soubiran}}{{Prugniel} et~al.}{2007}]{prugniel+07proc}
{Prugniel} P.,  {Koleva} M.,  {Ocvirk} P.,  {Le Borgne} D.,    {Soubiran} C.,
  2007, in {Vazdekis} A.,  {Peletier} R.~F.,  eds, Stellar Populations as
  Building Blocks of Galaxies Vol.~241 of IAU Symp, {Analysis of stellar
  populations with large empirical libraries at high spectral resolution}.
pp 68--72

\bibitem[\protect\citeauthoryear{{Prugniel} \& {Soubiran}}{{Prugniel} \&
  {Soubiran}}{2001}]{ELODIE}
{Prugniel} P.,  {Soubiran} C.,  2001, \aap, 369, 1048

\bibitem[\protect\citeauthoryear{{Prugniel}, {Soubiran}, {Koleva} \& {Le
  Borgne}}{{Prugniel} et~al.}{2007}]{ELODIE3}
{Prugniel} P.,  {Soubiran} C.,  {Koleva} M.,    {Le Borgne} D.,  2007, ArXiv
  Astrophysics e-prints

\bibitem[\protect\citeauthoryear{{Prugniel}, {Vauglin} \& {Koleva}}{{Prugniel}
  et~al.}{2011}]{prugniel+11}
{Prugniel} P.,  {Vauglin} I.,    {Koleva} M.,  2011, \aap, 531, A165

\bibitem[\protect\citeauthoryear{{Rajpurohit}, {Reyl{\'e}}, {Allard},
  {Homeier}, {Schultheis}, {Bessell} \& {Robin}}{{Rajpurohit}
  et~al.}{2013}]{rajpurohit+13}
{Rajpurohit} A.~S.,  {Reyl{\'e}} C.,  {Allard} F.,  {Homeier} D.,  {Schultheis}
  M.,  {Bessell} M.~S.,    {Robin} A.~C.,  2013, \aap, 556, A15

\bibitem[\protect\citeauthoryear{{Rodr{\'{\i}}guez-Merino}, {Chavez}, {Bertone}
  \& {Buzzoni}}{{Rodr{\'{\i}}guez-Merino} et~al.}{2005}]{UVBLUE05}
{Rodr{\'{\i}}guez-Merino} L.~H.,  {Chavez} M.,  {Bertone} E.,    {Buzzoni} A.,
  2005, ApJ, 626, 411

\bibitem[\protect\citeauthoryear{{S{\'a}nchez-Bl{\'a}zquez}, {Peletier},
  {Jim{\'e}nez-Vicente}, {Cardiel}, {Cenarro}, {Falc{\'o}n-Barroso}, {Gorgas},
  {Selam} \& {Vazdekis}}{{S{\'a}nchez-Bl{\'a}zquez} et~al.}{2006}]{MILES1}
{S{\'a}nchez-Bl{\'a}zquez} P.,  {Peletier} R.~F.,  {Jim{\'e}nez-Vicente} J.,
  {Cardiel} N.,  {Cenarro} A.~J.,  {Falc{\'o}n-Barroso} J.,  {Gorgas} J.,
  {Selam} S.,    {Vazdekis} A.,  2006, \mnras, 371, 703

\bibitem[\protect\citeauthoryear{{Sansom}, {de Castro Milone}, {Vazdekis} \&
  {S{\'a}nchez-Bl{\'a}zquez}}{{Sansom} et~al.}{2013}]{sansom+13}
{Sansom} A.~E.,  {de Castro Milone} A.,  {Vazdekis} A.,
  {S{\'a}nchez-Bl{\'a}zquez} P.,  2013, \mnras, 435, 952

\bibitem[\protect\citeauthoryear{{Sbordone}, {Bonifacio}, {Castelli} \&
  {Kurucz}}{{Sbordone} et~al.}{2004}]{ATLAS_LINUX}
{Sbordone} L.,  {Bonifacio} P.,  {Castelli} F.,    {Kurucz} R.~L.,  2004,
  Memorie della Societa Astronomica Italiana Supplement, 5, 93

\bibitem[\protect\citeauthoryear{{Schiavon}}{{Schiavon}}{2007}]{schiavon07}
{Schiavon} R.~P.,  2007, \apjs, 171, 146

\bibitem[\protect\citeauthoryear{{Schwenke}}{{Schwenke}}{1998}]{schwenke98}
{Schwenke} D.~W.,  1998, Chemistry and Physics of Molecules and Grains in
  Space.~ Faraday Discussions No.~109.~ The Faraday Division of the Royal
  Society of Chemistry, London, 1998., p.321, 109, 321

\bibitem[\protect\citeauthoryear{Shannon}{Shannon}{1998}]{shannon}
Shannon C.~E.,  1998, Proceedings of the IEEE, 86, 447

\bibitem[\protect\citeauthoryear{{Short} \& {Lester}}{{Short} \&
  {Lester}}{1996}]{short_lester96}
{Short} C.~I.,  {Lester} J.~B.,  1996, \apj, 469, 898

\bibitem[\protect\citeauthoryear{{Sordo}, {Vallenari}, {Tantalo}, {Allard},
  {Blomme}, {Bouret}, {Brott}, {Fremat}, {Martayan}, {Damerdji}, {Edvardsson},
  {Josselin} \& {Plez}}{{Sordo} et~al.}{2010}]{sordo+10}
{Sordo} R.,  {Vallenari} A.,  {Tantalo} R.,  {Allard} F.,  {Blomme} R.,
  {Bouret} J.-C.,  {Brott} I.,  {Fremat} Y.,  {Martayan} C.,  {Damerdji} Y.,
  {Edvardsson} B.,  {Josselin} E.,    {Plez} e.~a.,  2010, \apss, 328, 331

\bibitem[\protect\citeauthoryear{{Soubiran}, {Katz} \& {Cayrel}}{{Soubiran}
  et~al.}{1998}]{soubiran+98}
{Soubiran} C.,  {Katz} D.,    {Cayrel} R.,  1998, \aaps, 133, 221

\bibitem[\protect\citeauthoryear{{Strom} \& {Kurucz}}{{Strom} \&
  {Kurucz}}{1966}]{strom_kurucz66}
{Strom} S.~E.,  {Kurucz} R.,  1966, \aj, 71, 181

\bibitem[\protect\citeauthoryear{{Thomas}, {Maraston}, {Bender} \& {de
  Oliveira}}{{Thomas} et~al.}{2005}]{thomas+05}
{Thomas} D.,  {Maraston} C.,  {Bender} R.,    {de Oliveira} C.~M.,  2005, ApJ,
  621, 673

\bibitem[\protect\citeauthoryear{{Tody}}{{Tody}}{1986}]{IRAF1}
{Tody} D.,  1986, in {Crawford} D.~L.,  ed., Society of Photo-Optical
  Instrumentation Engineers (SPIE) Conference Series Vol.~627 of Society of
  Photo-Optical Instrumentation Engineers (SPIE) Conference Series, {The IRAF
  Data Reduction and Analysis System}.
p.~733

\bibitem[\protect\citeauthoryear{{Tody}}{{Tody}}{1993}]{IRAF2}
{Tody} D.,  1993, in {Hanisch} R.~J.,  {Brissenden} R.~J.~V.,   {Barnes} J.,
  eds, Astronomical Data Analysis Software and Systems II Vol.~52 of
  Astronomical Society of the Pacific Conference Series, {IRAF in the
  Nineties}.
p.~173

\bibitem[\protect\citeauthoryear{{Torres}, {Andersen} \&
  {Gim{\'e}nez}}{{Torres} et~al.}{2010}]{torres+10}
{Torres} G.,  {Andersen} J.,    {Gim{\'e}nez} A.,  2010, \aapr, 18, 67

\bibitem[\protect\citeauthoryear{{Trager}, {Worthey}, {Faber}, {Burstein} \&
  {Gonzalez}}{{Trager} et~al.}{1998}]{trager+98}
{Trager} S.~C.,  {Worthey} G.,  {Faber} S.~M.,  {Burstein} D.,    {Gonzalez}
  J.~J.,  1998, ApJS, 116, 1

\bibitem[\protect\citeauthoryear{{Tsuji}}{{Tsuji}}{1986}]{tsuji86}
{Tsuji} T.,  1986, \araa, 24, 89

\bibitem[\protect\citeauthoryear{{van Belle} \& {von Braun}}{{van Belle} \&
  {von Braun}}{2009}]{vanbelle+09}
{van Belle} G.~T.,  {von Braun} K.,  2009, \apj, 694, 1085

\bibitem[\protect\citeauthoryear{{Vazdekis}, {S{\'a}nchez-Bl{\'a}zquez},
  {Falc{\'o}n-Barroso}, {Cenarro}, {Beasley}, {Cardiel}, {Gorgas} \&
  {Peletier}}{{Vazdekis} et~al.}{2010}]{vazdekis+10}
{Vazdekis} A.,  {S{\'a}nchez-Bl{\'a}zquez} P.,  {Falc{\'o}n-Barroso} J.,
  {Cenarro} A.~J.,  {Beasley} M.~A.,  {Cardiel} N.,  {Gorgas} J.,    {Peletier}
  R.~F.,  2010, \mnras, 404, 1639

\bibitem[\protect\citeauthoryear{{Walcher}, {Coelho}, {Gallazzi} \&
  {Charlot}}{{Walcher} et~al.}{2009}]{walcher+09}
{Walcher} C.~J.,  {Coelho} P.,  {Gallazzi} A.,    {Charlot} S.,  2009, \mnras,
  398, L44

\bibitem[\protect\citeauthoryear{{Westera}, {Lejeune}, {Buser}, {Cuisinier} \&
  {Bruzual}}{{Westera} et~al.}{2002}]{basel3}
{Westera} P.,  {Lejeune} T.,  {Buser} R.,  {Cuisinier} F.,    {Bruzual} G.,
  2002, \aap, 381, 524

\bibitem[\protect\citeauthoryear{{Worthey}, {Faber} \& {Gonzalez}}{{Worthey}
  et~al.}{1992}]{worthey+92}
{Worthey} G.,  {Faber} S.~M.,    {Gonzalez} J.~J.,  1992, ApJ, 398, 69

\bibitem[\protect\citeauthoryear{{Worthey} \& {Lee}}{{Worthey} \&
  {Lee}}{2011}]{worthey_lee11}
{Worthey} G.,  {Lee} H.-c.,  2011, \apjs, 193, 1

\bibitem[\protect\citeauthoryear{{Zwitter}, {Castelli} \& {Munari}}{{Zwitter}
  et~al.}{2004}]{zwitter+04}
{Zwitter} T.,  {Castelli} F.,    {Munari} U.,  2004, \aap, 417, 1055

\end{thebibliography}


\end{document}